\pdfoutput=1

\documentclass[10pt,journal,compsoc]{IEEEtran}
\usepackage{amsmath,amssymb,amsfonts}
\usepackage{graphicx}
\usepackage{textcomp}
\usepackage{xcolor}
\usepackage{epstopdf}
\usepackage{algorithm}
\usepackage{algpseudocode}

\usepackage{pdfpages}
\usepackage{multirow}
\usepackage{xcolor}
\usepackage{todonotes}
\usepackage{caption}
\usepackage{subfigure}
\usepackage{amsthm}
\newtheorem{theorem}{Theorem}[section]
\usepackage{graphicx}
\usepackage{bm}
\usepackage{url}
\usepackage{threeparttable}
\newtheorem{defn}{Definition}[section] 
  
\usepackage[switch]{lineno}
\nolinenumbers

\ifCLASSOPTIONcompsoc
  \usepackage[nocompress]{cite}
\else
  \usepackage{cite}
\fi
\ifCLASSINFOpdf
\else
\fi
\begin{document}
%
\title{ICPE: An Item Cluster-Wise Pareto-Efficient Framework for Recommendation Debiasing}
%
%
%
%

\author{Yule~Wang, Xin~Xin, Yue Ding, Yunzhe Li, and Dong Wang
\IEEEcompsocitemizethanks{\IEEEcompsocthanksitem Yule Wang is with School of Computational Science and Engineering, Georgia Institute of Technology, Atlanta, GA 30332, USA.
E-mail: yulewang@gatech.edu.
\IEEEcompsocthanksitem Xin Xin is with School of Computer Science and Technology, Shandong University, Qingdao 266000, China. E-mail: xinxin@sdu.edu.cn.
\IEEEcompsocthanksitem Yue Ding and Dong Wang are with School of Electronic Information and Electrical Engineering, Shanghai Jiao Tong University, Shanghai 200240, China. E-mail: \{dingyue, wangdong\}@sjtu.edu.cn.
\IEEEcompsocthanksitem Yunzhe Li is with Department of Computer science, University of Illinois, Urbana-Champaign(UIUC), Champaign, IL 61820, USA. E-mail: yunzhel2@illinois.edu.
\IEEEcompsocthanksitem Corresponding Author: Dong Wang}
\thanks{Manuscript received October 16, 2022; revised October 25, 2023.}
}

%
%

\markboth{}%
{Shell \MakeLowercase{\textit{et al.}}: Bare Demo of IEEEtran.cls for Computer Society Journals}
%



\IEEEtitleabstractindextext{%
\begin{abstract}
Recommender system based on historical user-item interactions is of vital importance for web-based services. 
However, the observed data used to train the recommender model suffers from severe bias issues (e.g., exposure bias, popularity bias). 
Practically, the item frequency distribution of the dataset is a highly skewed power-law  distribution. Interactions of a small fraction of head (popular) items account for almost the whole training data. The normal training paradigm from such biased data tends to repetitively generate recommendations from the head items, which further exacerbates the biases and affects the exploration of potentially interesting items from the niche (long-tail) set. 
In this work, distinct from existing methods, we innovatively explore the central theme of recommendation debiasing from an item cluster-wise multi-objective optimization perspective. Through an empirical study, we find that head items are highly likely to be recommended because \emph{the gradients coming from head items dominate the overall gradient update process, which further affects the optimization of niche items}. Aiming to balance the learning on various item clusters that differ in popularity during the training process, we characterize the recommendation task with item cluster-wise Pareto-efficiency. To this end, we propose a model-agnostic framework namely \textit{\textbf{I}tem \textbf{C}luster-Wise \textbf{P}areto-\textbf{E}fficienct} Recommendation (ICPE). 
In detail, we define our item cluster-wise optimization target as the recommender model should balance all item clusters that differ in popularity, thus we set the model learning on each item cluster as a unique optimization objective. To achieve this goal, we first explore items' popularity levels from a novel causal reasoning perspective. Then, we devise popularity discrepancy-based bisecting clustering to separate the item clusters. Next, we adaptively find the overall harmonious gradient direction for multiple item cluster-wise optimization objectives from a Pareto-efficient solver. Finally, in the prediction stage, we perform counterfactual inference to further eliminate the impact of global propensity. We conduct experiments on three public datasets, instantiating ICPE with three state-of-the-art backbone recommender models. 
Extensive experimental results verify the superiorities of ICPE on overall recommendation performance and biases elimination. Codes will be open-source upon acceptance.
\end{abstract}

}

\maketitle

\IEEEdisplaynontitleabstractindextext

%
\IEEEpeerreviewmaketitle

\IEEEraisesectionheading{\section{Introduction} \label{introduction}}


Recommender systems (RS) play a crucial role in online services and platforms to address the problem of information overload \cite{he2017neural, shi2022task}. In conventional methods based on the most prevalent technique, Collabaritive Filtering (CF) \cite{he2020lightgcn,li2021extracting,chen2021airec}, a recommender model is trained using observed user-item historical feedbacks with the target of providing personalized recommendation items given the current user state. Since it's difficult to collect explicit feedback from various real-world applications, in this work we focus on implicit feedbacks, which are easily collected by storing users’ interaction behaviors (e.g., click), generating the dataset for further training steps.

\begin{figure}
  \centering
  \includegraphics[width=0.4\textwidth]{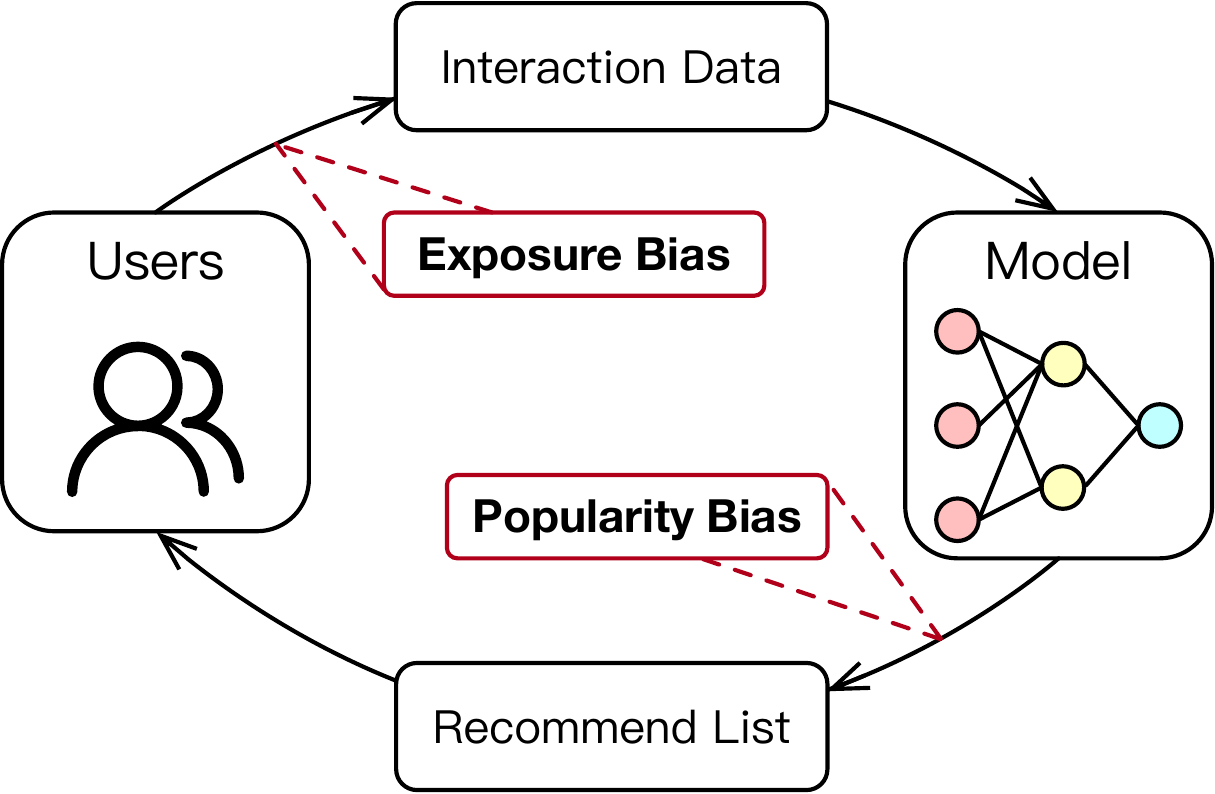}
  \caption{Biases Occurrence and Amplification Loop.}
  \label{self_loop}
\end{figure}

However, inherent bias issues emerge as great challenges for modern RS. Traditionally, the incremental training procedure of the recommender model in industrial applications of RS is a process of mutual dynamic evolution, where exists a severe bias amplification feedback loop \cite{chen2020bias}, as shown in Fig. \ref{self_loop}. Specifically, first, since a fraction of users are likely to interact with the head items, these actions will be collected by the RS. Then, as the head items are much more frequently treated as positive samples, they will be pushed towards much higher ranking scores compared with other items (i.e., popularity bias). After the RS generate the recommended list, the exposure mechanism of the RS results in the fact that users could only be exposed to the top head items and only interact with them (i.e., exposure bias), further affecting the collection of user-item interactions. As the circle back of training data and recommend list in the RS continues, the biases get amplified, introducing the notorious Matthew effect \cite{chen2020bias}.


Hence, debiasing in recommendation plays an important role in improving the system's performance. From the user's perspective, he/she could be easily bored with the head items repetitively recommended by the system. There are potentially relevant items that will lead to larger user satisfaction in the niche set but they have never been exposed \cite{wang2021deminet, adomavicius2011improving}.  
As for service providers, the recommendation from niche items can embrace more marginal profit compared with head items \cite{anderson2006long}. 
Generally speaking, the recommendation task is a typical exploitation-exploration problem. The model can exploit its known information and recommend items aligning with the users' interests, or it can recommend diversity items for identifying user's latent preferences. The debiased recommendation will benefit both users and service providers with better exploration, which finally turns into larger benefits in the long-run \cite{jannach2015recommenders}.

Several existing methods focused on recommendation debiasing are developed based purely on inverse propensity scoring (IPS) \cite{lee2022bilateral}, which re-weights the data samples with certain theoretical analyses. However, it introduces high variance in the implicit feedback scenario \cite{saito2020ubpr, swaminathan2015self}, the estimated weights fluctuate drastically in each step. Some other works introduce additional models \cite{chen2021autodebias, zhu2020unbiased} for pseudo-label or propensity score calculation, resulting in the estimation overlap issue \cite{ren2018learning}. On the other hand, methods using causal reasoning \cite{wei2021model, zhang2021causal} have explicitly modeled the impact of item/user popularity through causal graphs. However, they mainly rely on strong assumptions of the causal effects. Additionally, without rebalancing the data samples towards the unbiased distribution, the expressive power of causal methods is limited and they do not address the exposure bias in RS.

\begin{figure}
  \centering
  \subfigure[Gradient Norm Distribution]{\label{gradient}\includegraphics[width=0.24\textwidth]{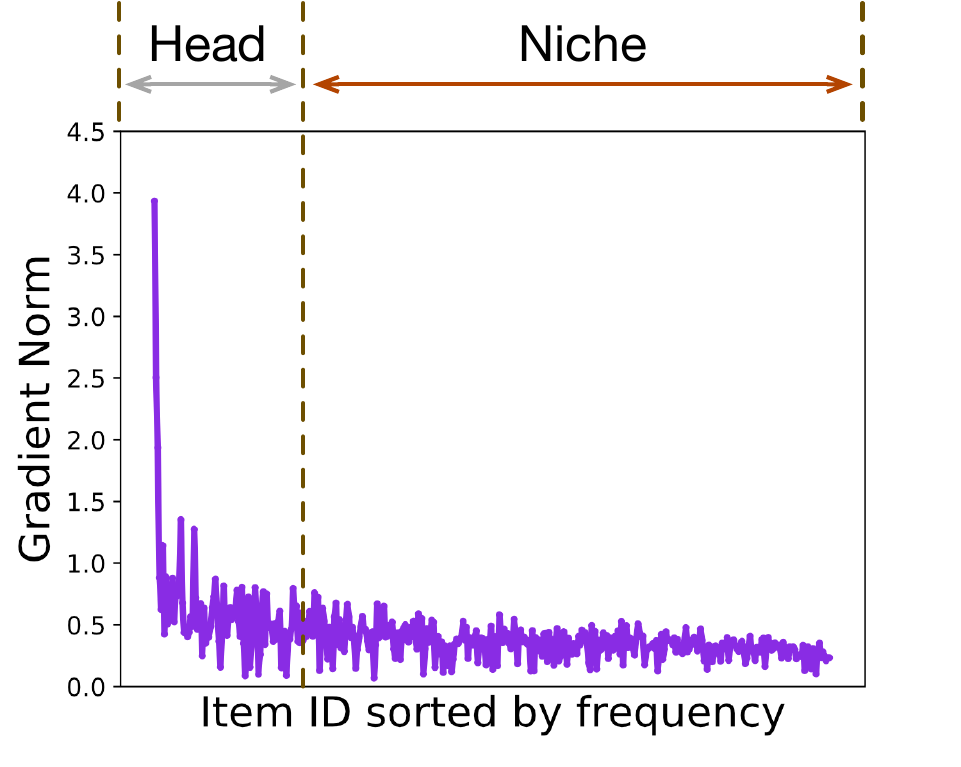}}
  \subfigure[Graident Conflict Example]
  {\label{conflict}\includegraphics[width=0.24\textwidth]{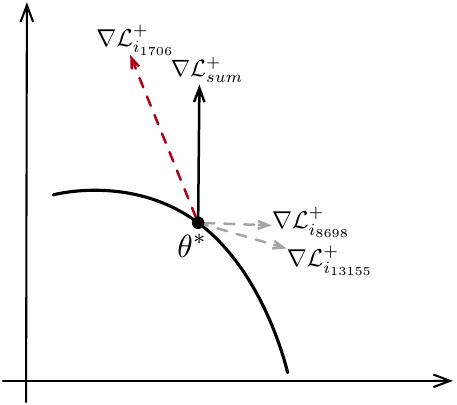}} 
  \vspace{-0.3cm}
\caption{Empirical study on Gowalla dataset.}
\vspace{-0.4cm}
\end{figure}
To eliminate the biases in a unified framework, in this article, we analyze the bias issues of recommender system from an optimization perspective. Specifically, we observe that the interaction frequencies of all the items in the training data follow an extreme power-law distribution \cite{AbdollahpouriBM17}. Next, we conduct an empirical study on the Gowalla\footnote{https://snap.stanford.edu/data/loc-gowalla.html} dataset, on which we train the state-of-the-art LightGCN \cite{he2020lightgcn} model. 
Fig. \ref{gradient} visualizes the norm (i.e., $L_2$-norm $||\cdot||_2$) of gradients coming from different items. 
Fig. \ref{conflict} gives examples of the overall gradients of one head item $i_{1706}$, two niche items $i_{8698}$, $i_{13155}$, and the total effect of the above three terms in one training epoch. We can draw the following observation:
\begin{itemize}
\item[---] Head items have much larger gradient norms than niche items, indicating that the overall gradient direction is dominated by head items. 
\item[---] There exists direction conflicts between gradients coming from head items and niche items. That is to say, updating model parameters based on gradients dominated by head items sacrifice the learning of niche items. 
\end{itemize}
Motivated by the above observation, we propose ICPE for recommendation debiasing. 
More precisely, we define our item cluster-wise optimization target that the recommender model should perform fairly for all item clusters that differ in popularity, setting the model learning on each item cluster as a unique optimization objective.  To achieve this goal, we first explore items' popularity levels from a novel causal reasoning perspective, calculating their correlation score with the user's global propensity. Then, we devise popularity discrepancy-based bisecting clustering to separate the discriminated item clusters, in which the feature and popularity information are both considered. Next, we adaptively find the weights of multiple item cluster-wise optimization objectives from a Pareto-efficient solver. As a result, the learning over the whole training data can be seen as a weighted aggregation of multiple cluster-wise objectives, reaching the overall harmonious gradient direction. Finally, in the prediction stage, we perform counterfactual inference to further eliminate the impact of global propensity.  To the best of our knowledge, this is the first attempt to explore  recommendation debiasing from an item cluster-wise multi-objective optimization perspective. To summarize, this work makes the following contributions:
\begin{itemize}
\item[---] 
Inspired by the gradient influence of various items to the recommender model, we propose to perform recommendation debiasing from an item cluster-wise multi-objective optimization perspective. In our approach, based on gradients, we characterize the recommendation task with Item Cluster-wise Pareto-Efficiency (ICPE) to balance their learning. 
\item[---] Focusing on recommendation debiasing, we propose a novel item clustering algorithm: Popularity Discrepancy-based Clustering. This algorithm leverages causal inference techniques to extract the global propensity and subsequently employs it to guide the bisecting AE-KMeans clustering.
\item[---] We have established theoretical guarantees to quantify the issue of gradient domination by head items in the normal training paradigm. And we also derive the generalization error bound guarantees for the proposed ICPE framework.
\end{itemize}

\section{PRELIMINARY}
\label{preliminary}

\subsection{Problem Formulation}
Let $\mathcal{U}=\left\{u_{1}, u_{2}, \ldots, u_{|\mathcal{U}|}\right\}$, and $\mathcal{I}=\left\{i_{1}, i_{2}, \ldots, i_{|\mathcal{I}|}\right\}$ denote the set of users and items, respectively, where $|\mathcal{U}|$ is the size of the user set, and $|\mathcal{I}|$ is the size of the item set. In this paper, we consider training the recommender model from implicit feedback (i.e., observed interactions are considered as positive samples while negative samples are sampled from missing interactions). The positive interaction set between the users and items is defined as: $\mathcal{R} = \{(u,i)|u \in \mathcal{U}, i \in \mathcal{I}, \text { user } u \text { interacted with item } i\}$. Formally, given the dataset of $\{\mathcal{U}, \mathcal{I}, \mathcal{R}\}$, our task is to learn a predictive recommender model, such that for each user, it can accurately rank all the candidate items in $\mathcal{I}$ according to his/her preference. After that, items with the Top-$N$ prediction scores will constitute the final recommendation list to the user.

However, various biases occur in real-world RS, leading to the inconsistency between the distribution of collected training data and the ideal unbiased data (we elaborate this phenomenon in Section \ref{sec:error}). Specifically, in this work, we will focus on eliminating the following two classes of bias shown in Fig. \ref{self_loop}:

\begin{itemize}
\item[---] \textbf{Exposure Bias:} \textit{This bias happens as users are constantly kept exposed to a small part of head items, so it's hard for the recommender model to learn their preference on the less observed niche item set.}
\item[---] \textbf{Popularity Bias:} \textit{The recommender model tends to over-estimate the ranking scores of popular items while under-estimating the ranking scores of niche items.}
\end{itemize}

\subsection{Bridging Item Popularity and Gradient Influence in Recommender Models}
\label{bridging}

As for a traditional recommender model extracting individual user interest as item representations, under the normal training setting (i.e., each training sample has equal weight), 
the total loss function is defined on the whole training data, which is shown as 
\begin{equation} 
\label{normal_sum_form}
\mathcal{L}^n = \mathcal{L}^+ + \mathcal{L}^- =  \sum_{(u,i)\in \mathcal{R}^+} \mathcal{L}_{u,i}+\sum_{(u,i)\in \mathcal{R}^-} \mathcal{L}_{u,i},
\end{equation}
where $\mathcal{R}^+$ and $\mathcal{R}^-$ denote the set of positive samples and the set of sampled negative samples, correspondingly, $\mathcal{L}_{u,i}$ is the specific loss (i.e., cross-entropy) of $(u,i)$ pair. Then we expand the loss on positive samples\footnote{We don't consider the loss on negative examples since we assume that they are sampled uniformly.} as 
\begin{equation} 
\label{eq:problem_form}
\mathcal{L}^+=\sum_{(u,i)\in \mathcal{R}^+}\ \mathcal{L}_{u,i}=\sum_{i\in \mathcal{I}}\sum_{u\in\mathcal{R}_i^+}\ \mathcal{L}_{u,i} = \sum_{i\in \mathcal{I}} \mathcal{L}_{i}^{+},
\end{equation}
where $\mathcal{I}$ denotes the whole item set and $\mathcal{R}_i^+$ is the set of users who have interacted with item $i$ on positive samples. 

In Eq. (\ref{eq:problem_form}), due to the popularity bias in training data, $\mathcal{R}_i^+$ is extremely imbalanced. For head items, $\mathcal{R}_i^+$ contains much more samples than niche items. As a result, given shared parameter $\mathbf{w}_{sh}$, when we perform updates according to $\partial \mathcal{L}_{i}^{+}/\partial \mathbf{w}_{sh}$, the majority of gradients would come from the loss on head items. When there are conflicts between gradients coming from head items and gradients coming from niche items, the normal training setting would scarify the learning on niche items to achieve a lower overall loss. That is to say, the overall gradient direction is dominated by head items.

\section{ICPE: PROPOSED METHODOLOGY}
\label{methodology}





\begin{figure*}[t]
  \centering
  \includegraphics[width=0.95\textwidth]{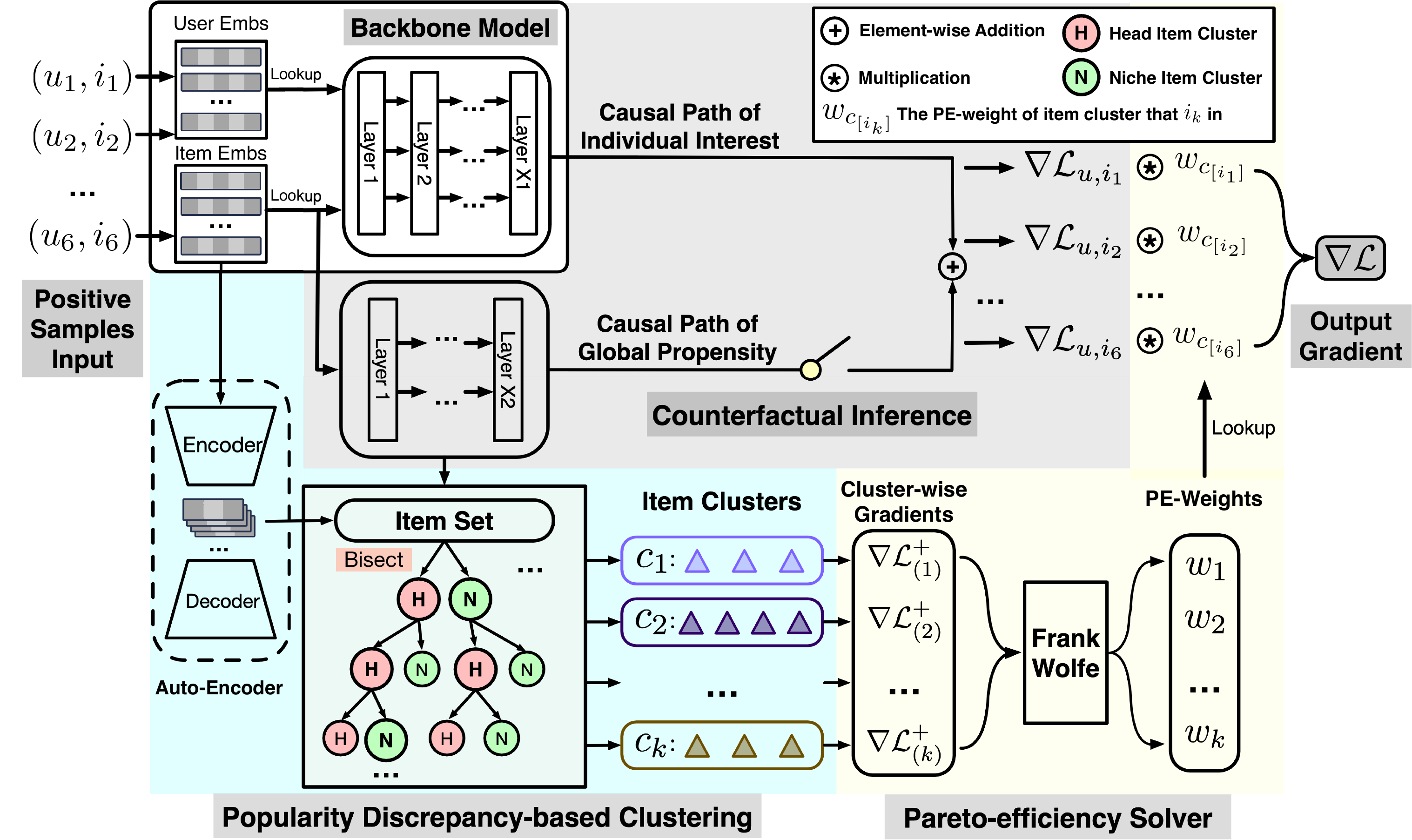}
  \caption{The schematic overview of ICPE. Firstly, according to the training target that the recommender model should balance all clusters of items that differ in popularity, we explore items' actual popularity level (via a specific neural model) from a novel cause-effect perspective and then devise Popularity Discrepancy-based Clustering to separate the discriminated item clusters. Next, we acquire the non-conflict gradient direction for multiple item cluster-wise objectives from the Pareto-efficiency solver. Finally, we combine the above components in the training stage and perform counterfactual inference in the prediction stage to further eliminate the impact of global propensity. }
  \label{framework}
\end{figure*}

To tackle the two challenges discussed in Section \ref{introduction} and perform popularity debiasing for recommender systems, we propose to cluster the items with discrepancies in popularity and treat them as multiple optimization objectives, resulting in our ICPE framework. Specifically, it consists of three aspects: \textit{popularity discrepancy-based clustering}, \textit{Pareto-efficiency solver}, and \textit{counterfactual inference}. The workflow of ICPE is provided in Fig. \ref{framework}. In this section, we will detail each component after the illustration of 
\textit{Item Cluster-Wise Pareto-Efficiency Principle}. 

\subsection{Item Cluster-Wise Pareto-Efficiency Principle}


Our method aims to find a set of weights for the item-wise objectives based on their gradients, so that they can be optimized without hurting the others.
However, in practice, the size of the item set $|\mathcal{I}|$ can be extremely large. As a result, naively considering each item-wise loss $\mathcal{L}_i^+$ as an optimization objective mainly suffers from two problems: (1) the high variance of the reweighted loss\cite{swaminathan2015self,wan2022cross}; (2) the unacceptable computational cost of finding the weights. To address these problems, ICPE divides the whole item set into $K$ clusters ($K << |\mathcal{I}|$) and treats the training target as an item cluster-wise multi-objective optimization problem. 

Formally, in the total parameter learning space $\bm{\theta}$ of recommender model, there exist certain parameters $\bm{\theta}^{sh}$ which are shared between item cluster-wise objectives and other parameters $\bm{\theta}^{os}$ which are only correlated to a specific item cluster objective. ICPE focuses on the solution of objective weights on $\bm{\theta}^{sh}$, since $\bm{\theta}^{os}$ would not affect the learning of other item clusters. 
Taking all the above factors, we formulate the loss function of ICPE to train $\bm{\theta}^{sh}$ on positive samples as
\begin{equation} 
\begin{aligned}
\label{sum_form} 
    \mathcal{L}^{sh+} & = \sum_{k=1}^{K} w_k \mathcal{L}_{(k)}^+\text{, where } \mathcal{L}_{(k)}^+=\sum_{i\in \mathcal{N}_k} \mathcal{L}_i^+,\\
	\text{s.t.}&  \ \  \sum_{k=1}^{K} w_k = 1, w_k \geq h_k, \ \text{for} \ \ k = 1,...,K 
\end{aligned}
\end{equation}
where $w_k$ is the weight for the $k$-th item cluster $c_k$, $\mathcal{N}_k$ denotes the set of items in $c_k$, and $\{h_1 ,..., h_K \}$ are the debiasing boundary constraints for the $K$ objectives. With proper constraint settings on each cluster-wise objective, ICPE pushes the weight solution to the middle part of the Pareto Frontier. When $K=1$, we can recover the normal training weight setting.
$\mathcal{L}_{(k)}$ is the learning objective for cluster $k$.
For the training of objective-specific parameters $\bm{\theta}^{os}$, we still use $\mathcal{L}^+$ as shown in Eq. (\ref{normal_sum_form}).



\subsection{Popularity Discrepancy-based Clustering}
\label{subsec:PDC}

\subsubsection{Global Propensity of the User Set}

For item clustering, rustically clustering the items through the absolute ranking of interacted times is overly empirical. Here we first explore a new method that measures the level of popularity discrepancy between item subsets and set it as a clustering criterion. More specifically, for item subset popularity level measurement, we calculate its contained items' correlations with \textit{global propensity, which denotes the overall preference trend of the entire user set}. To capture the global propensity, we introduce a parameterized neural network with casual science that learns from the data samples. Leveraging the derived global propensity, we then design a novel clustering algorithm.

\begin{figure}[t]
\centering    
\subfigure[Causal graph of ICPE that considers global propensity] {
\label{ICPE_training}     
\label{ICPE_training}     
\includegraphics[width=0.21\textwidth]{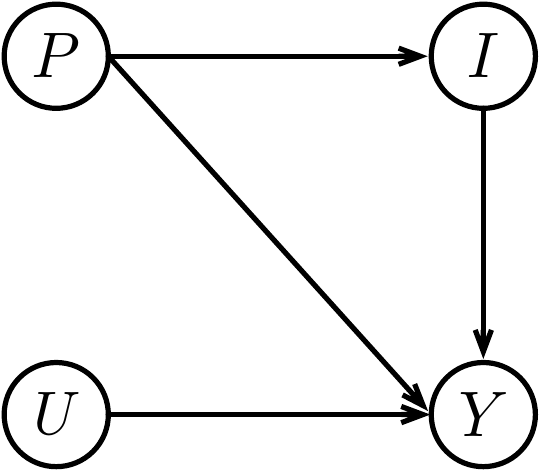}  
}
\subfigure[Counterfactual Inference of ICPE] {
\label{ICPE_inference}     
\includegraphics[width=0.21\textwidth]{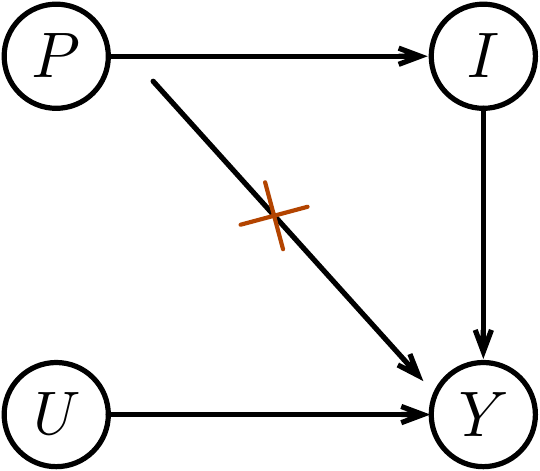}  
}
\caption{Causal graphs describing the recommender model. $U$: user, $I$: candidate item, $P$: global propensity, $Y$: interaction probability. We identify $P$ as a confounder between $I$ and $Y$, and propose counterfactual inference by reducing the natural direct effect from $P$ on $Y$.}
\label{pareto} 
\vspace{-0.35cm}
\end{figure}



Considering the coarse-grained causal graph of traditional methods, i.e., $\hat{y}_{u,i}=f(u, i)$. Different motivations of user are bundled into a single causal path, leading to bias issues and inferior robustness \cite{scholkopf2022causality}. As discussed above, global propensity plays a key role in interaction probability due to the ``herd mentality of users'', which means they tend to follow the majority to consume popular items. Thus in ICPE, we explicitly consider how global propensity affects the interaction process, enriching the causal graph to Fig. \ref{ICPE_training}. That is to say, we decompose each user interaction into two disentangled factors, \textit{global propensity} and \textit{individual interest}. Node $P$ denotes global propensity, which has causal effects on $I$ and $Y$. We now provide the structural causal model (SCM) of our proposed framework:
\begin{align}
\label{ii_scm} & S_{u,i}^{n} :=f_{1}\left(U=u, I=i\right) = \text{BACKBONE}\left(\mathbf{x}_{u}, \mathbf{x}_{i} \right) ,\\
\label{gp_scm} & S_{p,i}^{g} :=f_{2}\left(I=i, P=\boldsymbol{\theta}_p\right) = \text{MLP}_{\boldsymbol{\theta}_p}\left(\mathbf{x}_{i} \right),\\
\label{y_hat_scm} & \hat{Y}_{u,i} :=f_{3}\left(S_{u,i}^{n}, S_{p,i}^{g} \right) = \left(1 - \alpha_g\right) S_{u,i}^{n} + \alpha_g S_{p,i}^{g}.
\end{align}

where $\mathbf{x}_{u}, \mathbf{x}_{i}$ are the embeddings of user and candidate item from the backbone model; and $f_{1}(\cdot), f_{2}(\cdot)$ and $f_{3}(\cdot)$ are the underlying causal mechanisms for item's excitation of the user individual interest, item's correlation with the global propensity, and interaction probability, respectively; $S^{n} \in \mathbb{R}^{\left|\mathcal{U}\right| \times\left|\mathcal{I}\right|}, S^{g} \in \mathbb{R}^{\left|\mathcal{I}\right|}$, and $\hat{Y}_{u,i} \in \mathbb{R}^{\left|\mathcal{U}\right| \times\left|\mathcal{I}\right|}$ represent the corresponding prediction scores of the above three causal mechanisms. In practice, $f_{1}(\cdot)$ is set as the backbone model; we parameterize $f_{2}(\cdot)$ with a neural network in which $LeakyReLU (\cdot)$ is set as the non-linear activation function and the global propensity of the user set is represented by its learnable parameters $\boldsymbol{\theta}_p$; under the mild assumption that the interaction probability can be factorized into global propensity and individual interest, we set $f_{3}(\cdot)$ as a concise additive causal model, in which $\alpha_g$ is a positive hyper-parameter to further control the weights of $S_{u,i}^{n}$ and $S_{p,i}^{g}$. 

Recently, many works design novel backbone models, $f_1(\cdot)$, to use structural information of the data \cite{he2020lightgcn} or learn non-linear feature interaction functions \cite{he2017neural}. However, the impact of global propensity on the candidate item is ignored by these methods. In the above causal framework, we can calculate each item's correlation with the global propensity through $f_{2}(\cdot)$.

\subsubsection{Adaptive Item Clustering}

Na\"ively clustering the items according to their occurrences neglects properties of the dataset and remains static during the training stage. To enhance the debiasing ability, we aim that for items with similar features, they are recommended fairly by the model, rather than be affected by their popularity. Hence, semantics of items should be considered in clustering so that items similar in feature but vary in popularity should be assigned into different clusters (i.e., optimization objectives). 

To this end, we propose an unsupervised Popularity Discrepancy-based Bisecting Clustering algorithm (detailed in Algorithm \ref{GPH_KM}). Firstly, in each iteration $t$, the algorithm performs a bisection on the cluster with the largest variance of score $S^{g}$ calculated in Eq. (\ref{gp_scm}) by the Bisecting AE-KMeans \cite{wang2017research}, in which an autoencoder is introduced for high-level feature extraction from the original latent space of items representation $\mathbf{X}_{ \mathcal{I}}$. We denote the fetched original cluster as $c_t$ and the bisected new clusters as $c_{t_1}$ and $c_{t_2}$.

\textit{Popularity Discrepancy Measurement.} Aiming to find item clusters differ in popularity, we define a metric $\tilde{D}$ to measure the level of popularity discrepancy of $c_t$, $c_{t_1}$ and $c_{t_2}$, which is calculated by:
\begin{equation} 
\begin{aligned}
\label{POP_BIAS} 
& \tilde{D}_{(\mathcal{N}_k)} = \left\|\overline{S^{g}}_{(\mathcal{N}_k)} - \overline{S^{g}}_{(\mathcal{I}\backslash\mathcal{N}_k)}\right\|_{1}, \\
&\text{where } \overline{S^{g}}_{(\mathcal{N})} = \frac{1}{|\mathcal{N}|} \sum_{i \in \mathcal{N}} f_{2}(\mathbf{x}_{i}), \text{ given item set  } \mathcal{N}.
\end{aligned}
\end{equation}

In the above equations, $f_{2}(\cdot)$ is the learned underlying causal mechanism of global propensity in Eq. (\ref{gp_scm}); given cluster $c_k$, $\overline{S^{g}}_{(\mathcal{N}_k)}$ and $\overline{S^{g}}_{(\mathcal{N}\backslash\mathcal{N}_k)}$ refer to the average correlation score with the global propensity of items in and out of $c_k$, respectively; $\tilde{D}_{(\mathcal{N}_k)}$ is the L1-norm of the difference between the prior two terms, measuring the level of popularity discrepancy between items in and out of $c_k$. If the popularity discrepancy of any split new cluster $\{c_{t_{1}}, c_{t_{2}}\}$ is greater than that of the original cluster $c_t$, the split is performed; otherwise, $c_t$ is kept. This bisection step repeats until the desired number of clusters $K$ is reached.

\begin{algorithm}
 \caption{Popularity Discrepancy-based Clustering}
  \label{GPH_KM}
 	\begin{algorithmic}[1]
        \renewcommand{\algorithmicrequire}{\textbf{Input:}} 
        \renewcommand{\algorithmicensure}{\textbf{Output:}}
        \Require
        $\text{Maximum cluster number } K;$ \  $\text{Item set } \mathcal{I} \text{ with embeddings } \mathbf{X}_{ \mathcal{I}} = \{\mathbf{x}_1, \mathbf{x}_2,\ldots, \mathbf{x}_{i}\}$
        \Ensure
        $\text{Item clusters } \mathcal{C} $
        \State Initialize item cluster: $c_0 \leftarrow \mathcal{I}$
        \State $h \leftarrow max\_heap([(\sigma^{2}_{c_0}, c_0)])$ {\Comment{$\sigma^{2}:$ the key of $h$, denoting cluster's variance of $S^{g}$ in Eq. (\ref{gp_scm})}}
        \While{$h.size() < K$ \textbf{and} \textit{not} $h.isEmpty()$}
        \State $(\sigma^{2}_{c_t}, c_t) \leftarrow h.peek()$  {\Comment{Cluster $c_t$ with the largest variance of $S^{g}$}}
        \State Perform Bisecting AE-KMeans on $c_t$, get $c_{t_1}, c_{t_2}$
       \If {max$\left(\tilde{D}_{(\mathcal{N}_{t_1})},\tilde{D}_{(\mathcal{N}_{t_2})}\right) \geq \tilde{D}_{(\mathcal{N}_{t})}$}
           \State $h.pop()$
           \State $h.push((\sigma^{2}_{c_{t_1}}, c_{t_1}),(\sigma^{2}_{c_{t_2}}, c_{t_2}))$
       \EndIf
       \EndWhile
    \State \Return $h.clusters$
 	\end{algorithmic}
 \end{algorithm}

\subsection{Pareto-efficient Solver}
Obtaining the item clusters $\mathcal{C}$, in the following, we will describe how to find $w_k$ for each cluster objective. Firstly, we provide a brief introduction to Pareto-efficiency and some related concepts.

Given a model aiming to minimize a series of objective functions $\mathcal{L}_{(1)}^+, . . . , \mathcal{L}_{(K)}^+ $, Pareto-efficiency is a state when it is impossible to improve one objective without hurting other objectives. 
Formally, we provide the following definition:

\begin{defn}
For a minimization task of multiple objectives, let $s_m$ and $s_n$ denote two solutions as $s_m = (f _1^m , . . . , f_K^m )$ and $s_n = (f _1^n , . . . , f_K^n )$, $s_m$ \textbf{dominates} $s_n$ if and only if $f_1^m \leq f_1^n$, . . . , $f_K^m \leq f_K^n$.
\end{defn}
Then the concept of Pareto-efficiency is defined as:

\begin{defn}
\label{Pareto-efficient}
Solution $s = (f_1 , . . . , f_K )$ is \textbf{Pareto-efficient} if and only if there is no other solution that dominates $s$ .
\end{defn}

In ICPE, we aim to find $w_k$ so that the solution of each cluster-wise objective is \emph{Pareto-efficient}. Formally, according to Definition \ref{Pareto-efficient}, we use the Karush-Kuhn-Tucker (KKT) conditions \cite{wu2007karush} to describe the property and constraints of finding $w_k$:
%
\begin{equation}
\begin{aligned}
\label{transformed}
	\mathop{min}_{w_1, \ldots, w_K} &  \left \|  \sum_{k=1}^{K} w_k \nabla_{\bm{\theta}^{sh}} \mathcal{L}_{(k)}^+ \right\|^2 _2, \\ 
	\text{$s.t.$} \ \  &  \sum_{k=1}^{K} w_k = 1, w_k \geq h_k, \ \text{for} \ \ k = 1,...,K. 
\end{aligned}
\end{equation}
The quadratic programming problem defined in (\ref{transformed}) is equivalent to finding a minimum-norm point in the convex hull of the set of input points \cite{makimoto1994efficient}. We obtain a set of Pareto-efficient solutions by employing the Frank-Wolfe solver \cite{jaggi2013revisiting} to (\ref{transformed}) multiple times. We denote the total number of runs as $T$. Thus, we get a set of Pareto-efficient solutions $\mathbf{W} = \{\mathbf{w}_1, \mathbf{w}_2, \ldots, \mathbf{w}_T \}$.

As ICPE is a generic framework for recommendation debiasing, we use `fairness' as the principal criterion for solution selection. Specifically, we utilize the intuitive and widely-adopted \cite{xiao2017fairness, agarwal2017does} `Least Misery' \cite {basu2015group} method. To align with the focus on mitigating the gradient domination issue of head items, the `miserable' metric we propose here is selecting the solution that minimizes the largest $L^2$ norm (Euclidean norm) within the gradients of $K$ cluster-wise objectives:

\begin{equation}
\begin{aligned}
& \mathbf{w}^* = \underset{\mathbf{w} \in \mathbf{W}}{\operatorname{argmin}} \max \left\{g_1, g_2, \ldots, g_K\right\}, \\
\text{ where \:} &  g_k = w_k \left\| \nabla_{\boldsymbol{\theta}^{sh}} \mathcal{L}_{(k)}^{+}\right\|_2, \text{ for } k = 1, \ldots, K.
\end{aligned}
\end{equation}

\subsection{Optimization and Counterfactual Inference}

\subsubsection{Weighted Binary Cross-Entropy}

We use binary cross-entropy (BCE) as the basement loss to optimize parameters. More precisely, the specific loss $\mathcal{L}_{u,i}$ for a $(u,i)$ pair is formulated as: 
\begin{equation}
\mathcal{L}_{u,i} = - y_{u,i} \log(\sigma(\hat y_{u,i})) - (1 -y_{u,i}) \log(1- \sigma(\hat y_{u,i})),
\end{equation}
where $y_{u,i}$ is the ground-truth label for the $(u,i)$ pair, $\sigma(\cdot)$ is the sigmoid function. $\hat y_{u,i}$ is acquired according to Eq. (\ref{y_hat_scm}). The optimization function of ICPE regarding the shared parameter $\bm{\theta}^{sh}$ is formulated as: 
\begin{equation}
\label{L_ICPE_sh}
\mathcal{L}^{sh} = \mathcal{L}^{sh+} + \mathcal{L}^{-} + \lambda_r\| \bm{\theta}^{sh} \|^2_2,
\end{equation}
where $\mathcal{L}^{sh+}$ is calculated according to Eq. (\ref{sum_form}), $\mathcal{L}^-$ is calculated according to Eq. (\ref{normal_sum_form}), $\lambda_r$ is regularization coefficient. For objective-specific parameter $\bm{\theta}^{os}$, we maintain the normal setting, the optimization function is formulated as:
\begin{equation}
\label{L_ICPE_os}
\mathcal{L}^{os} = \mathcal{L}^{n} + \lambda_r \| \bm{\theta}^{os} \|^2_2,
\end{equation}
where $\mathcal{L}^n$ is calculated according to Eq. (\ref{normal_sum_form}).

Algorithm \ref{training_alg} illustrates the overall training procedure of ICPE. In each training batch, as shown in line $5-7$, ICPE first calculates the score function of the user individual interest causal path and global propensity causal path, which are combined into the total prediction score. Next, the items are clustered through the Popularity Discrepancy-based Clustering in line $8$. ICPE then performs the PE-solver to obtain the weight $w_k$ for each item cluster $c_k$ in line $9$. While in line $10-13$ we calculate the loss functions for shared and objective-specific parameters, then perform the update.

\begin{algorithm}
    \caption{The Training Procedure of ICPE}
    \label{training_alg}
    \begin{algorithmic}[1] 
    \renewcommand{\algorithmicrequire}{\textbf{Input:}} 
    \renewcommand{\algorithmicensure}{\textbf{Output:}}
    \Require
     Dataset $\mathcal{R^+}$,$\mathcal{R^-}$; Backbone recommender model $\mathcal{M}$; Maximum item cluster number $K$; Learning rate $\eta$ and all other hyperparameters;
    \Ensure
    Parameters in both the shared and objective-specific learning space $\bm{\theta}$: \{$\bm{\theta}^{sh}$, $\bm{\theta}^{os}\}$;
    \State Initialize $\bm{\theta}$;
    \While {not converge}
    \State $(u,i) \leftarrow$ Sample a mini-batch from $\mathcal{R^+}$ and $\mathcal{R^-}$;
    \State $\mathbf{x}_{u}, \mathbf{x}_{i} \leftarrow$ Lookup user, item representations from $\mathcal{M}$;
    \State $S_{u,i}^{n} \leftarrow$ Calculate score function of user individual interest causal path using Eq. (\ref{ii_scm});
    \State $S_{p,i}^{g} \leftarrow$ Calculate score function of global propensity causal path using Eq. (\ref{gp_scm});
    \State $\hat{Y}_{u,i} \leftarrow$ Calculate total prediction score function based on $S_{u,i}^{n}$ and $S_{p,i}^{g}$ using Eq. (\ref{y_hat_scm});
    \State $\mathcal{C} :=\left\{c_{1}, c_{2}, \ldots, c_{K}\right\} \leftarrow$ Generate $K$ item clusters using Algorithm \ref{GPH_KM};
    \State $\bm{w} :=\left\{w_1, w_2, \ldots, w_K\right\} \leftarrow$ Update cluster weights using Frank-Wolfe Solver; 
    \State $\mathcal{L}^{sh} \leftarrow$ Compute loss for shared parameters based on $\hat{Y}_{u,i}$ and $\bm{w}$ according to Eq. (\ref{L_ICPE_sh}); 
    \State $\mathcal{L}^{os} \leftarrow$ Compute loss for objective-specific parameters based on $\hat{Y}_{u,i}$ according to Eq. (\ref{L_ICPE_os}); 
    \State $\bm{\theta}^{sh} \leftarrow \bm{\theta}^{sh} - \eta \cdot \partial \mathcal{L}^{sh}/\partial \bm{\theta}^{sh} $;
    \State $\bm{\theta}^{os} \leftarrow \bm{\theta}^{os} - \eta  \cdot \partial \mathcal{L}^{os}/\partial \bm{\theta}^{os} $;
    \EndWhile
    \State \Return $\bm{\theta}$
\end{algorithmic}
\end{algorithm}

\subsubsection{Counterfactual Inference}

Considering the causal graph in Fig. \ref{ICPE_training}, node $P$ (denoting global propensity) has two edges pointing to node $I$ and node $Y$, respectively. The path $P \rightarrow I$ refers to the better intrinsic quality of the item given its high popularity, while path $P \rightarrow Y$ means the global propensity directly impacts the user behaviors, which increases the exposure of head items even though they may not match users' true interest. For answering the counterfactual question: \textit{``what the prediction score would be if the framework only considers user individual interest"}, we remove the bad effect of popularity bias on model inference through cutting off the path $P \rightarrow Y$ (shown in Fig. \ref{ICPE_inference}), making the prediction scores free from the global propensity we modeled in Eq. (\ref{gp_scm}). To this end, we perform counterfactual inference as follows: 
\begin{equation} 
\begin{aligned}
\label{ICPE_Inference_Func} 
\tilde{Y}_{u,i} := S^{n}_{u,i},
\end{aligned}
\end{equation}

where $\tilde{Y}_{u,i}$ is the final prediction score. Algorithm \ref{inference_alg} describes the procedure of inference in ICPE.

\begin{algorithm}
    \caption{Inference of ICPE}
    \label{inference_alg}
    \begin{algorithmic}[1] 
    \renewcommand{\algorithmicrequire}{\textbf{Input:}} 
    \renewcommand{\algorithmicensure}{\textbf{Output:}}
    \Require
    User $u$; Item $i$; Backbone recommender model $\mathcal{M}$;
    \Ensure
    Prediction score $\tilde{y}_{u,i}$
    \State $\mathbf{x}_{u}, \mathbf{x}_{i} \leftarrow$ Lookup representations of $u$, $i$ from $\mathcal{M}$;
    \State $S_{u,i}^{n} \leftarrow$ Calculate score function of user individual interest causal path using Eq. (\ref{ii_scm});
    \State $\tilde{y}_{u,i} \leftarrow$ Assign final score function directly based on $S_{u,i}^{n}$ according to Eq. (\ref{ICPE_Inference_Func});
    \State \Return $\tilde{y}_{u,i}$
\end{algorithmic}
\end{algorithm}




\subsection{Complexity Analyses of ICPE}
For the space complexity, ICPE just introduces a neural network for modeling the item’s correlation with the global propensity in Eq. (\ref{gp_scm}). Compared to the large number of item and user embeddings in the parameter set, the additional space cost of ICPE is negligible in practice.

In the following part, we will analyze the time complexity of ICPE. Compared to the normal training setting of the backbone model, the training process of ICPE contains two additional terms in each step: Item Clustering and PE-Solver. Suppose the embedding size of the backbone model is $d$, by iterating $K$ times, the total complexity of the item clustering process takes 
$\mathcal{O}\left(K * |\mathcal{I}| * d\right)$; For the PE-Solver in Algorithm $2$, its complexity mainly depends on the matrix multiplications in line $2$, the dimension of the shared parameters mainly comes from the user embeddings, whose size is $|\mathcal{U}| * d$. Therefore, this part costs $\mathcal{O}\left(K^2 * |\mathcal{U}| * d\right)$. As a result, the additional time complexity of ICPE is $\mathcal{O}\left(K * |\mathcal{I}| * d + K^2 * |\mathcal{U}| * d\right)$.

It should be noticed that in the inference stage of ICPE, since the rating scores are calculated from the user individual interest causal path alone (in Eq. \ref{ii_scm}), the time complexity of inference keeps the same as the backbone model. 



 \section{THEORETICAL ANALYSIS}
\label{theory}

\subsection{Power growth of Model Gradient Influence given Item Popularity}
To extract the correlation between overall gradient and item populrity, we introduce ideas from influence functions \cite{koh2017understanding} to measure the gradient influence of a specific item on the shared parameters $\bm{\theta}^{sh}$ of the recommender model. We first define gradient influence function $G(\cdot)$ from item feature $\mathbf{x}_{i}$ to one vector parameter $\mathbf{w}_{k}$ in the shared parameters set $\bm{\theta}^{sh}$ of the backbone recommender model as follows:
\begin{equation}
G(\mathbf{x}_{i}, \mathbf{w}_{k})=\left\| \nabla_{\mathbf{w}_{k}}  \mathcal{L}_{i}^+ \right\|_2,
\label{eq:grad_inf}
\end{equation}
We can draw to the relation of gradient influence and item popularity in the following theorem:


\begin{theorem}
\textit{Assume binary cross-entropy is the target loss function and Sigmoid is the activation function of the final interaction layer. As $\mathcal{R}_i^+$ is the set of users who have interacted with item $i$ in positive samples, given one vector parameter $\mathbf{w}_{k}$, the gradient influence $G(\mathbf{x}_{i}, \mathbf{w}_{k})$ is a power function, in which $|\mathcal{R}_i^+|$ is the function's base: $G(\mathbf{x}_{i}, \mathbf{w}_{k})  \ \propto \ |\mathcal{R}_i^+|^{(1 + \alpha_{I})} $. The value of $\alpha_{I}$ is determined by the recommender model architecture.}
\end{theorem}

\begin{proof}
We first expand the right hand side in Eq. (\ref{eq:grad_inf}) as:

\begin{equation}
\left\| \nabla_{\mathbf{w}_{k}}  \mathcal{L}_{i}^+ \right\| _2  =
\left\| \sum_{u \in \mathcal{R}_i^+}  
\nabla_{\mathbf{w}_{k}} \mathcal{L}_{u,i} \right\|_2 .
\end{equation}
We denote $\sigma(\cdot)$ as the \textit{Sigmoid} activation function and $f_{I}(\cdot)$ as the model interaction function (from which the logits are obtained). Thus we yield:
\begin{align}
\sum_{u \in \mathcal{R}_i^+} &  \nabla_{\mathbf{w}_{k}} \mathcal{L}_{u,i} = \sum_{u \in \mathcal{R}_i^+} \frac{\partial \mathcal{L}_{u,i}}{\partial f_{I}(\mathbf{x}_{u},\mathbf{x}_{i})} \cdot \nabla_{\mathbf{w}_{k}} f_{I}(\mathbf{x}_{u},\mathbf{x}_{i}) \\ & = \sum_{u \in \mathcal{R}_i^+} \left[\mathbf{1} - \sigma(f_{I}(\mathbf{x}_{u},\mathbf{x}_{i}))\right] \nabla_{\mathbf{w}_{k}} f_{I}(\mathbf{x}_{u},\mathbf{x}_{i}) \\
& \approx \mu(\mathbf{x}_{i}) \cdot \left( \sum_{u \in \mathcal{R}_i^+} \nabla_{\mathbf{w}_{k}} f_{I}(\mathbf{x}_{u},\mathbf{x}_{i}) \right) \ \propto \ |\mathcal{R}_i^+|^{(1 + \alpha_{I})} \label{eq:appro}. 
\end{align}
We draw to (\ref{eq:appro}) since for an untrained recommender model, $\mathbf{x}_{u}$ and $\mathbf{x}_{i}$ are normally distributed \cite{datta2020survey}, in which $0 < \mu(\mathbf{x}_{i}) < 1$ represents an expectation term correlatting with $\mathbf{x}_{i}$. Additionally, we extend the Node Degree Bias proposed in \cite{tang2020investigating} (\textit{Theorem 4.1}) to a general recommender model case to acquire the proportional relation in (\ref{eq:appro}). $\alpha_{I} = 0$ when there exists no graph modeling in $f_{I}(\cdot)$ (e.g., PMF, NeuMF), $\alpha_{I} \geq \frac{1}{2}$ if $f_{I}(\cdot)$ considers the graph message passing mechanism (e.g., LGC). 
\end{proof}



\subsection{Generalization Error Bound of ICPE}
\label{sec:error}
We further conduct theoretical analysis to verify that the distribution learnt by the ICPE framework is consistent with the ideal unbiased data distribution. We provide the generalization error bound of ICPE by modeling the bias issues in recommendation as a distribution shift problem. 

Given the pareto-efficient weights $\bm{w}$, we first denote the expectation risk of the optimum model parameters on the ideal unbiased distribution as $\mathcal{R}_{\bm{w}}(\bm{\theta}^{*})$. On the collected training data, we denote the empirical model parameters learnt by ICPE as $\widehat{\mathcal{R}}_{\bm{w}}(\bm{\theta})$. Since ICPE focuses on a multi-objective optimization objective, we resort to Rademacher complexity \cite{cortes2020agnostic,hoffman2018algorithms} to derive the error bound guarantee. With a mixture of the weights $\bm{w}$, the Rademacher complexity of $\bm{\theta}$ is as follows:
\begin{equation}
	\widehat{\mathcal{R}}_S(\bm{\theta}, \bm{w})=\underset{\boldsymbol{\sigma}}{\mathbb{E}}\left[\sum_{k=1}^K \frac{w_k }{|c_k|} \sum_{i=1}^{|c_k|} \sigma_i \mathcal{L}_i^+(\bm{\theta})\right],
\end{equation}

in which $\boldsymbol{\sigma}$ follows a uniform distribution between $[-1,1]$.
We assume that $\forall (\mathbf{x}_i, \mathbf{x}_u,y_{u,i}), (\mathbf{x}^{\prime}_i, \mathbf{x}^{\prime}_u,y^{\prime}_{u,i}) \in \mathcal{U} \times \mathcal{I} \times \mathcal{Y}, \forall \ \bm{\theta} \in \bm{\Theta}:\left\|\left(f_{I}(\mathbf{x}^{\prime}_{u},\mathbf{x}^{\prime}_{i}), y^{\prime}_{u,i}\right)-\left(f_{I}(\mathbf{x}_{u},\mathbf{x}_{i}), y_{u,i}\right)\right\| \leq \mathcal{D}_{\bm{\Theta}}$. Through the use of $\widehat{\mathcal{R}}_S(\bm{\theta}, \bm{w})$, we have the following theorem that bridges the gap between the expectation and empirical risks $\mathcal{R}_{\bm{w}}(\bm{\theta}^{*})$ and $\widehat{\mathcal{R}}_{\bm{w}}(\bm{\theta})$:

\begin{theorem}
\textit{If the loss functions $\mathcal{L}_{(k)}^+(\bm{\theta})$ are $M_k$-Lipschitz and bounded by $M$, then for any $\delta>0$, with probability at least $1-\delta$, the following inequality holds for $\forall \bm{\theta} \in \bm{\Theta}$ and $\forall \bm{w} \in \bm{W}$}:
\label{th:second_theorem}
\end{theorem}
\begin{equation}
\begin{aligned}
	\mathcal{R}_{\bm{w}}(\bm{\theta}^{*}) \leq \widehat{\mathcal{R}}_{\bm{w}}(\bm{\theta}) + & \widehat{\mathcal{R}}_S(\bm{\theta}, \bm{w}) \\ + & \mathcal{D}_{\Theta} \sum_{k=1}^K w_k M_k \sqrt{\frac{2}{T} \ln \left[\frac{|\mathbf{W}|_\epsilon}{\delta}\right]},
\end{aligned}
\label{eq:bound}
\end{equation} 

where $T = |\mathcal{R}| * K$, $|\mathbf{W}|_\epsilon$ is a minimum $\epsilon$-cover of $\mathbf{W}$.

\begin{proof}
	Due to the limited space, we briefly provide the main steps of the proof here. We first expand the discrepancy between $\mathcal{R}_{\bm{w}}(\bm{\theta}^{*})$ and $\widehat{\mathcal{R}}_{\bm{w}}(\bm{\theta})$ through McDiarmid's inequality, the result is composed of two terms: $\mathbb{E}\left[\sup _{\bm{\theta} \in \bm{\Theta}} \mathcal{R}_{\bm{w}}(\bm{\theta}^{*})-\widehat{\mathcal{R}}_{\bm{w}}(\bm{\theta})\right] = \Delta$, and the third term on the right-hand-side of inequality (\ref{eq:bound}). Then, using the definition of the minimum $\epsilon$-cover, we have $\Delta \leq \mathcal{R}_{\bm{w}}(\bm{\theta}^{*})$. Combining these two conclusions, Theorem \ref{th:second_theorem} is proofed.
\end{proof}

From Theorem \ref{th:second_theorem}, we conclude that when enough training data are collected, ICPE can find a nearly optimal debiasing parameter solution, leading the recommender model to approximate the unbiased optimum. We can also observe that the bound of ICPE can be tightened with the increasement of item cluster number $K$. However, the numerical solutions from the Pareto-efficiency conditional Solver would be unstable if we impose a large $K$. Thus there exists a trade-off between the setting of item cluster number.

 
\section{Experimental Setup}

\label{experiment}

%
%
%
%



\subsection{Experimental Settings}

\subsubsection{Datasets and Pre-processing} 
We conduct experiments on three public accessible datasets: Last.Fm \footnote{\url{https://files.grouplens.org/datasets/hetrec2011/hetrec2011-lastfm-2k.zip}},  Gowalla and Yelp2018 \footnote{\url{https://www.kaggle.com/yelp-dataset/yelp-dataset/version/7}}. The datasets vary in scale, domain, and sparsity. 

\textbf{Last.Fm}: This is a widely used dataset that contains $1$ million ratings between users and movies. We binarize the ratings into implicit feedback. Interacted items are considered as positive samples. Due to the sparsity of the dataset, we use the 10-core setting.
 
 \textbf{Gowalla}: This is the check-in dataset obtained from Gowalla, where users share their locations by checking-in behavior \cite{liang2016modeling}. To ensure the quality of the dataset, we use the 20-core setting.
 
 \textbf{Yelp2018}: This dataset is adopted from the 2018 edition of the Yelp challenge. Wherein, the local business shops like restaurants and bars are viewed as the items. Similarly, we use the 20-core setting to ensure that each user and item have at least 20 interactions.

All the samples in the above datasets are binarized into interacted or not. To reflect the average recommendation performance over the item set and get rid of the power-law distribution of the raw dataset \cite{AbdollahpouriBM17}, we need a debiased test set for a fair evaluation. To this end, we follow the data preprocessing methods of previous approaches \cite{zheng2021disentangling,wei2021model} to construct an unbiased test set where the interactions are sampled from a uniform distribution over items (all the samples will be re-weighted by the debiasing algorithms). 

\subsubsection{Evaluation Protocols} 
We adopt cross-validation to evaluate the performance. 
The ratio of training / validation / test set is 80\% / 10\% / 10\% and we split them chronologically. Each experiment is repeated 5 times and the average performance is reported. The recommendation quality is measured both in terms of overall accuracy and debiasing ability. 

The overall accuracy is measured with two metrics:  Recall@N and Normalized Discounted Cumulative Gain@N (NDCG@N). Recall@N measures how many ground-truth items are included in the top-N positions of the recommendation list. 
NDCG@N is a rank-sensitive metric that assigns higher weights to top positions in the recommended list.

For the evaluation of recommendation debiasing ability, we first split the whole item set $\mathcal{I}$ into $\mathcal{I}_h$ and $\mathcal{I}_n$ according to the ratio of 20\% / 80\% \textit{Pareto Principle}. $\mathcal{I}_h$ represents the set of head items and $\mathcal{I}_n$ denotes the set of niche items. 
Here 20\% means the top 20\% of the total item numbers, other than 20\% of the total interactions. 
We then adopt the following six metrics. 

\textbf{Recall-Head@N} and \textbf{NDCG-Head@N}: Recall-Head@N measures how many head items belong to $\mathcal{I}_h$
	are included in the top-N positions of the recommendation list and interacted with the user. Similarly, NDCG-Head@N only considers head items.

\textbf{Recall-Niche@N} and \textbf{NDCG-Niche@N}:Recall-Niche@N measures how many niche items belong to $\mathcal{I}_n$
	are included in the top-N positions of the recommendation list and interacted with the user. Similarly, NDCG-Niche@N only considers niche items.

\textbf{Coverage@N} and \textbf{APT@N}: Coverage measures the coverage of all the top-N recommendation lists to the whole item set $\mathcal{I}$. Average Percentage of niche items (APT) among the recommendation list is another more readily interpretable but closely related metric we use for evaluation. Precisely, Coverage@N and APT@N are defined as:
\begin{equation}
    \begin{aligned}
    Coverage@N & = \frac{|\cup_{u\in test} list_{@N}(u)|}{|\mathcal{I}|} 
    \end{aligned}
\end{equation}

\begin{equation}
    \begin{aligned}
    APT@N & = \frac{1}{|\mathcal{U}|} \sum_{u \in test} \frac{| list_{@N}(u) \cap \mathcal{I}_n |}{|list_{@N}(u)|} \\ 
    \end{aligned}
\end{equation}

where $list_N(u)$ represents the list of top-N recommended items for each user $u$ in the test set.

\begin{table}
\setlength\tabcolsep{3pt} 
\begin{center}
	\large
    \caption{Top-20 overall recommendation accuracy on three datasets. R@20 and NG@20 are short for Recall@20 and NDCG@20, respectively. Boldface denotes the highest score.}
  \label{table_overall}

  \resizebox{0.50\textwidth}{0.24\textwidth} 
  {%
  \begin{tabular}{cccccccc}
    \hline 
    
    \multicolumn{2}{c}{} & \multicolumn{2}{c}{Last.Fm} & \multicolumn{2}{c}{Gowalla} & \multicolumn{2}{c}{Yelp2018}\\
	\hline
    Backbone & Method
    & R@20 & NG@20 & R@20 & NG@20  & R@20 & NG@20 \\
    \hline
    \multirow{8}*{MF}&Normal & 0.0353 &	0.0343 & 0.1406 & 	0.1319 &  0.0382 &	0.0291   \\
    &DICE & 0.0366 &	0.0352 &0.1614 &0.1398 & 0.0456 &	0.0345     \\
    &MACR & 0.0381 &	0.0362 & 0.1639 &	0.1406 & 0.0480 &	0.0360      \\
    &CPR & 0.0383 &	0.0369 &0.1719 &	0.1463 & 0.0507 &	0.0377   \\
    &CausPref & 0.0379 &	0.0362 &0.1632 &	0.1452 & 0.0455 &	0.0341   \\
    &BC-Loss & 0.0377 &	0.0367 &0.1650 &	0.1451 & 0.0481 &	0.0370   \\
    &InvPref & 0.0395 &	0.0375 &0.1715 &	0.1463 & 0.0508 &	0.0381   \\
    &\textbf{ICPE} & \textbf{0.0409*}  & \textbf{0.0387*}   & \textbf{0.1754*}  & \textbf{0.1489*}  & \textbf{0.0513*}  & \textbf{0.0386*}    \\
    \hline
    \multirow{8}*{NeuMF}&Normal & 0.0323 & 	0.0309& 0.1245 &	0.1181  &0.0362 &	0.0279  \\
    &DICE & 0.0330 &	0.0318 & 0.1416 &	0.1300 & 0.0395 & 	0.0294      \\
    &MACR &	0.0348 &	0.0326 & 0.1440 	&0.1329 & 0.0439 &	0.0339    \\
    &CPR &0.0356 &	0.0329 & 0.1498 & 	0.1330 & 0.0448 &	0.0340    \\
    &CausPref & 0.0353 &	\textbf{0.0349} &0.1516 &	0.1325 & 0.0448 &	0.0348   \\
    &BC-Loss & 0.0350 &	0.0331 &0.1525 &	0.1323 & 0.0423 &	0.0344   \\
    &InvPref & 0.0336 &	0.0308 &0.1530 &	0.1342 & 0.0408 &	0.0333   \\
    &\textbf{ICPE} & \textbf{0.0367*}  &  0.0341  & \textbf{0.1587*}  & \textbf{0.1363*}  & \textbf{0.0476*}  &  \textbf{0.0357*}   \\
\hline
    \multirow{8}*{LGC}&Normal & 0.0396  &	0.0383 & 0.1594 &	0.1378 & 0.0423  	&0.0312   \\
    &DICE & 0.0407  &	0.0382 &0.1669 &	0.1439 &0.0514 &	0.0369     \\
    &MACR & 0.0413  &	0.0391 & 0.1762 &	0.1544 & 0.0546 &	0.0415     \\
    &CPR &0.0421  &	0.0399 & 0.1873 & 0.1616 & 0.0580 &	0.0437    \\
    &CausPref & 0.0399 &	0.0396 &0.1717 &	0.1483 & 0.0486 &	0.0369   \\
    &BC-Loss & 0.0380 &	0.0360 &0.1667 &	0.1444 & 0.0504 &	0.0435   \\
    &InvPref & 0.0407 &	0.0382 &0.1741 &	0.1585 & 0.0581 &	\textbf{0.0460}   \\
    &\textbf{ICPE} & \textbf{0.0432*} &	\textbf{0.0404*}  &	\textbf{0.1898 *} &	\textbf{0.1626*} &	\textbf{0.0599*} 	& 0.0448  \\
    \hline
  \end{tabular}}
  \begin{tablenotes}\footnotesize
    \item $*$ denotes significance p-value \textless 0.02 compared with the best baseline.
\end{tablenotes}
\end{center}
\end{table}

\subsubsection{Baselines}
We instantiate the proposed ICPE with three renowned recommender models, including classic matrix factorization, neural-network-based model, and graph-based model: 

\begin{itemize}
	\item[---] \textbf{Matrix Factorization (MF)} \cite{hu2008collaborative}: The most widely-used recommender model using linear user and item embeddings for production-based prediction score.
	\item[---] \textbf{Neural Matrix Factorization (NeuMF)} \cite{he2017neural}: NeuMF is one notable deep learning-based recommender model. It combines matrix factorization and multi-layer perceptrons (MLP) to learn high-order interaction signals.
	\item[---] \textbf{LightGCN} \cite{he2020lightgcn}: LightGCN is a state-of-the-art graph-based model that learns user and item representations by linearly propagating them on the interaction graph. The user and item embedding is formulated as 
	the aggregation of hidden vectors in all layers.

\end{itemize}
Each model is trained with the following model-agnostic debiasing frameworks:
\begin{itemize}
	\item[---] \textbf{Normal Training (Normal)}: The normal training procedure shown in Eq. (1).
	\item[---] \textbf{Disentangling Interest and Conformity with Causal Embedding (DICE)} \cite{zheng2021disentangling}: This is a state-of-the-art debiasing method that learns disentangled embeddings for the two causes, user interest and conformity.
	\item[---] \textbf{Model-Agnostic Counterfactual Reasoning(MACR)} \cite{wei2021model}: It is a causal-reasoning method that utilizes counterfactual inference to remove the direct effect of item properties on rank scores, mitigating the popularity bias.
	\item[---] \textbf{Cross Pairwise Ranking (CPR)} \cite{wan2022cross}: It is a method forming the loss term as the combination of multiple observed interactions at once, removing the influence of data biases caused by the exposure mechanism.
	\item[---] \textbf{CausPref: Causal Preference Learning for Out-of-Distribution Recommendation} \cite{wang2023extraction}: This is a novel causal preference-based framework that incorporates recommendation-specific Directed Acyclic Graph (DAG) learning, focusing on invariant user preference and anti-preference negative sampling.
	\item[---] \textbf{Bias-aware margins into Contrastive loss (BC-Loss)} \cite{zhang2022incorporating}: This method aims to mitigate the negative impact of popularity bias on collaborative filtering models by incorporating bias-aware margins tailored to the bias degree of each user-item interaction.
	\item[---] \textbf{Invariant Preference Learning (InvPref)} \cite{wang2025exploring}: This is a general debiasing framework that leverages environment inference to exploit latent heterogeneity in data and captures invariant user preferences to address both known and unknown biases.
	\item[---] \textbf{ICPE}: Our proposed learning framework.
\end{itemize}

\subsubsection{Parameter Settings} 
All methods are learned with the Adam optimizer \cite{kingma2014adam} except using RMSprop optimizer in NeuMF-based models. 
The batch size is set as 1024. 
The learning rate is set as $1e-3$. 
We evaluate the validation set every 3000 batches of updates. 
For a fair comparison, the embedding size is set as 64 for all models. 
For NeuMF and LightGCN, we utilize a three-layer structure.
The node-dropout and message-dropout in LightGCN are set as $0.1$ on all datasets.
For hyperparameters of ICPE, the parameters \(\alpha\) and \(\lambda_r\) are varied within the set \(\{1 \times 10^{-4}, 1 \times 10^{-3}, 2 \times 10^{-3}, 5 \times 10^{-3}, 1 \times 10^{-2}\}\) across all three datasets. The maximum item cluster number $K$ is searched between $\{3,4,5,6\}$ on various datasets. 


\subsection{Performance Comparison (RQ1)}

Experimental results show the superiority of ICPE in terms of both accuracy and debiasing ability.

\subsubsection{Overall Accuracy Comparison}

Table \ref{table_overall} shows the overall accuracy performance of top-N recommendation on all three datasets: Last.Fm, Gowalla, and Yelp2018, respectively. Although ICPE is proposed to tackle the bias issues in RS, in all cases, it outperforms normal training and other debiasing frameworks in terms of overall accuracy. It demonstrates that ICPE achieves a better trade-off between debiasing ability and overall accuracy compared with other frameworks. The performance gain regarding the overall accuracy metrics NDCG@20 / Recall@20 is 4.03 \% / 2.98 \%. This improvement mainly comes from promoting high-quality niche items while downgrading irrelevant head items.
 
\subsubsection{Debiasing Capability Comparison} 
\begin{figure*}
\centering    
\subfigure[Last.fm] {
\includegraphics[width=0.3\textwidth]{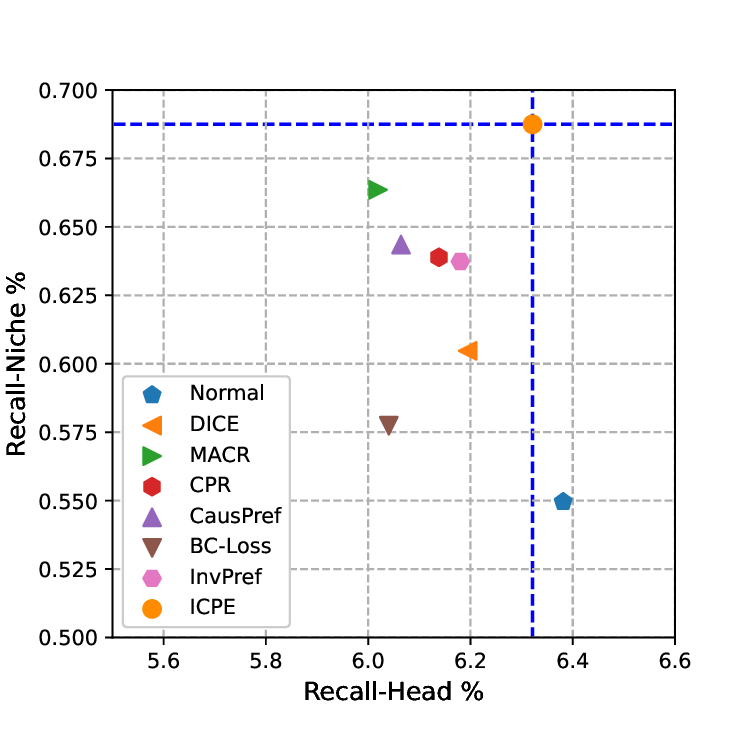}
} 
\subfigure[Gowalla] {
\includegraphics[width=0.3\textwidth]{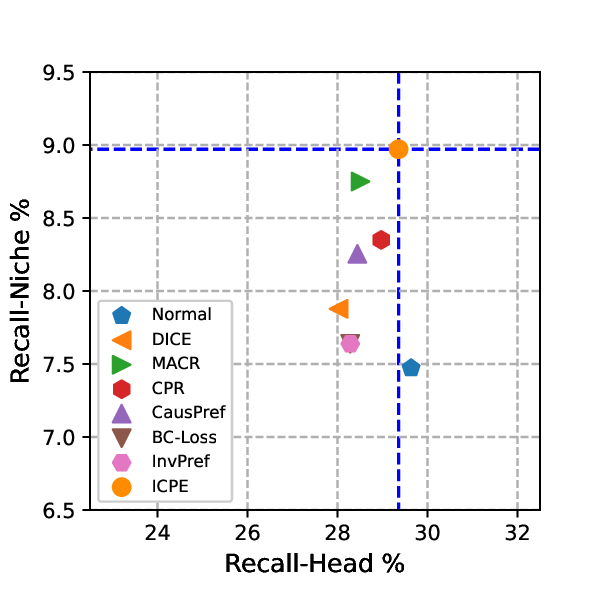}  
}
\subfigure[Yelp2018] {
\includegraphics[width=0.3\textwidth]{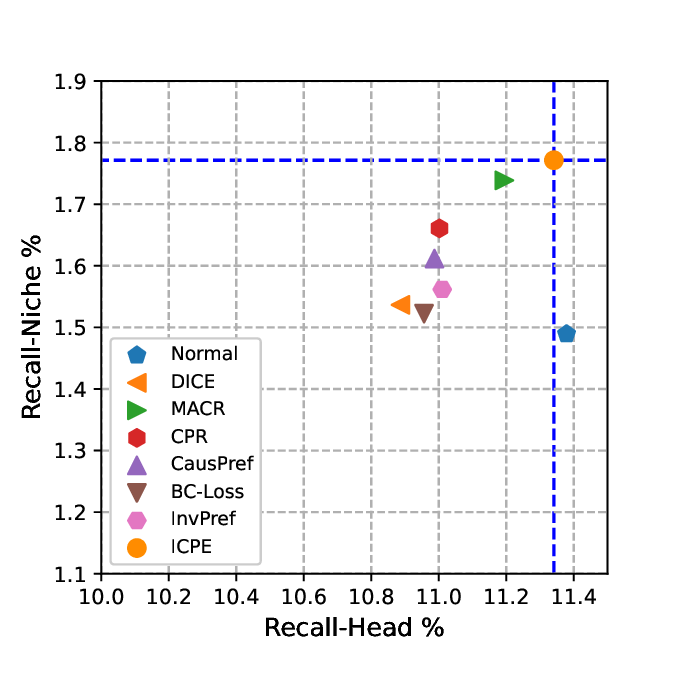}  
}
\caption{Recall-Head@20 and Recall-Niche@20 on LightGCN}     
\label{trade_off_recall} 
\end{figure*}

\begin{figure*}
\centering    
\vspace{-0.5cm}
\subfigure[Last.fm] {
\includegraphics[width=0.3\textwidth]{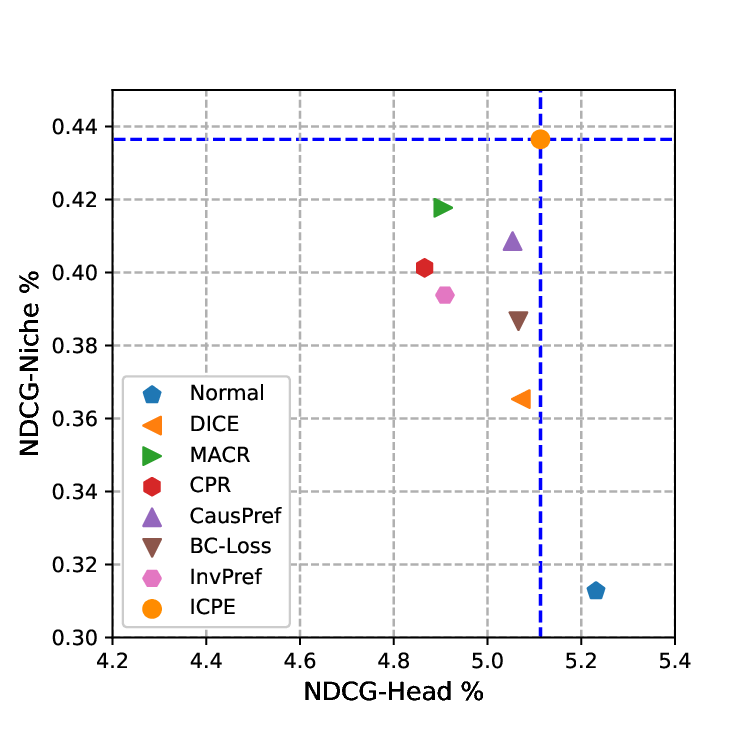}  
} 
\subfigure[Gowalla] {
\includegraphics[width=0.3\textwidth]{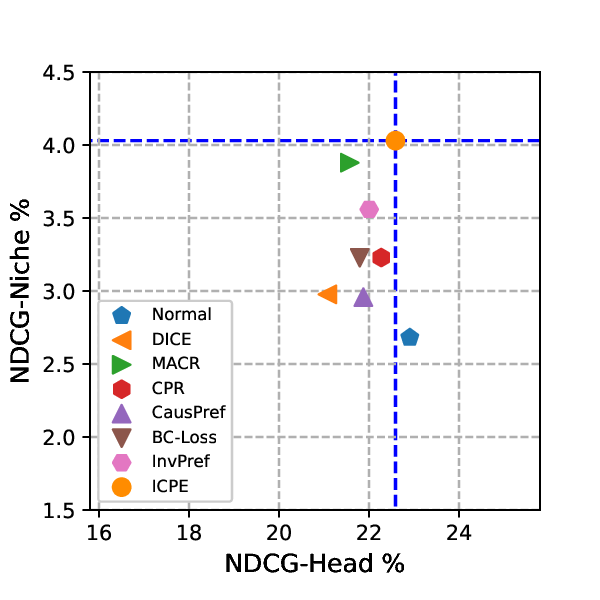}  
}
\subfigure[Yelp2018] {
\includegraphics[width=0.3\textwidth]{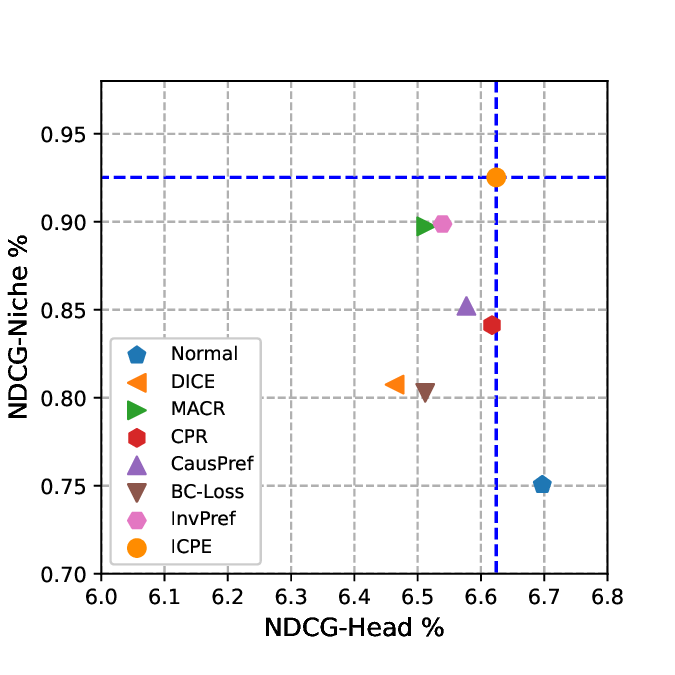}  
}
\caption{NDCG-Head@20 and NDCG-Niche@20 on LightGCN}   
\label{trade_off_ndcg} 
\end{figure*}

Fig. \ref{trade_off_recall} and Fig. \ref{trade_off_ndcg} visualize the performance of ICPE and other competing methods in terms of Recall-Head@20, Recall-Niche@20, NDCG-Head@20, and NDCG-Niche@20 on the state-of-the-art LightGCN backbone model. A similar trend can also be observed in MF and NeuMF. Table \ref{table_unbiasedness} shows the performance of ICPE and other competing methods in terms of Coverage@20 and APT@20. We have the following observations:

(1). According to Table \ref{table_unbiasedness}, ICPE achieves the best debiasing ability among all methods. 
This observation confirms that the proposed ICPE is effective to alleviate the bias issues and generate better recommendations on the niche item set. Meanwhile, the item coverage rate gets greatly improved, demonstrating the fairness of ICPE. Specifically, the backbone model with ICPE achieves the average absolute Coverage@20 / APT@20 gain of 15.45\% / 33.23\%.

(2). In Fig. \ref{trade_off_recall} and Fig. \ref{trade_off_ndcg}, Generally, the points which fall into the top-right area indicate better performance with both higher head and niche accuracy. 
It's obvious that the proposed ICPE achieves the highest niche accuracy while only scarifies a little head accuracy compared with normal training. 
However, other methods can not achieve such a performance. 
In most cases, these methods tend to lead to a larger decrease in head accuracy while obtaining a smaller gain in niche accuracy. 
This result demonstrates that ICPE achieves a better balance between head items and niche items, compared with other methods. The performance gain of ICPE mainly comes from the growth in niche items without the loss across head items.


(3). We conduct one-sample t-tests and the obtained results (i.e., p-value
\textless 0.01) indicate that the improvement regarding both debiasing ability metrics and overall accuracy metrics of ICPE is statistically significant.

\begin{table}
\begin{center}
	\large
    \renewcommand\tabcolsep{2.0pt}
  \caption{Top-20 recommendation debiasing capability on three datasets. Cov@20 is short for Coverage@20. AT@20 is short for APT@20. Boldface denotes the highest score.}
  \label{table_unbiasedness}
  \vspace{-0.2cm}

  \resizebox{0.50\textwidth}{0.24\textwidth}
  {%
  \begin{tabular}{cccccccc}
    \hline 
    \multicolumn{2}{c}{} & \multicolumn{2}{c}{Last.Fm} & \multicolumn{2}{c}{Gowalla} & \multicolumn{2}{c}{Yelp2018}\\
    Backbone & Method
    & Cov@20 & AT@20 & Cov@20 & AT@20  & Cov@20 & AT@20 \\
    \hline
    \multirow{8}*{MF}&Normal & 0.3034 &	0.0713 & 0.3064 & 	0.0926 &  0.2845 &	0.0377   \\
    &DICE & 0.3692 &	0.1252 &0.3531 &0.1233 & 0.3286 &	0.0345     \\
    &MACR & 0.3914 &	0.1373 & 0.3640 &	0.1259 & 0.3437 &	0.0360      \\
    &CPR & 0.3761 &	0.1214 &0.3722 &	0.1332 & 0.3512 &	0.0377   \\
    &CausPref & 0.4007 &	0.1401 &0.3511 &	0.1373 & 0.3452 &	0.0357   \\
    &BC-Loss & 0.3856 &	0.1317 &0.3342 &	0.1318 & 0.3653 &	0.0345   \\
    &InvPref & 0.4218 &	0.1467 &0.3420 &	0.1398 & 0.3525 &	0.0329   \\
    &\textbf{ICPE} & \textbf{0.4785*}  & \textbf{0.1542*}   & \textbf{0.4112*}  & \textbf{0.1451*}  & \textbf{0.3818*}  & \textbf{0.0386*}    \\
    \hline
    \multirow{8}*{NeuMF}&Normal & 0.5580 & 	0.1778 & 0.5053 &	0.1531  &0.4740 &	0.0695  \\
    &DICE & 0.6235 &	0.2336 & 0.5908 &	0.2166 & 0.5617 & 	0.0969      \\
    &MACR &	0.6188 &	0.2447 & 0.6377 	&0.2589 & 0.5875 &	0.1081    \\
    &CPR &0.6348 &	0.2644 & 0.6581 & 	0.2650 & 0.5897 &	0.1089    \\
    &CausPref & \textbf{0.6627*} &	\textbf{0.2832*} &0.6346 &	0.2501 & 0.5616 &	0.0963   \\
    &BC-Loss & 0.6026 &	0.2369 &0.6345 &	0.2629 & 0.5469 &	0.0943   \\
    &InvPref & 0.6264 &	0.2430 &0.6034 &	0.2366 & 0.5936 &	0.1184   \\
    &\textbf{ICPE} & 0.6418  &  0.2770  & \textbf{0.6724*}  & \textbf{0.3188*}  & \textbf{0.6350*}  &  \textbf{0.1339*}   \\
\hline
    \multirow{8}*{LGC}&Normal & 0.3508  &	0.1065 & 0.3547 &	0.0920 & 0.2955   	&0.0545   \\
    &DICE & 0.4115  &	0.1790 &0.3912 &	0.1190 &0.3767 &	0.0708     \\
    &MACR & 0.4211  &	0.1851 & 0.3710 &	0.1391 & 0.3841 &	0.0923     \\
    &CPR &0.4273  &	0.1835 & 0.4307 & 0.1331 & 0.3923 &	0.0919    \\
    &CausPref & 0.4131 &	0.1565 &0.4597 &	0.1454 & 0.4182 &	0.1027   \\
    &BC-Loss & 0.3805 &	0.1715 &0.4310 &	0.1445 & 0.3949 &	0.0925   \\
    &InvPref & 0.4108 &	0.1833 &0.4407 &	0.1458 & 0.4239 &	0.1100   \\
    &\textbf{ICPE} & \textbf{0.4418*} &	\textbf{0.2043*}  &	\textbf{0.4860*} &	\textbf{0.1600*} &	\textbf{0.4677*} 	& \textbf{0.1168*}  \\
    \hline
  \end{tabular}}
  \begin{tablenotes}\footnotesize
    \item $*$ denotes significance p-value \textless 0.01 compared with the best baseline.
\end{tablenotes}
\vspace{-0.45cm}
\end{center}
\end{table}


\subsection{Ablation Study and Hyper-parameter Study (RQ2)}

\subsubsection{Ablation Study}
In this part, we conduct an ablation study to analyze the functionality of the main components of ICPE (i.e., item cluster-wise re-weighting (CR), popularity discrepancy-based bisecting clustering (PD), and counterfactual inference (CI)).
Table \ref{ablation} shows the performance of ICPE and its variants on all three datasets using LightGCN as the backbone recommender model. We introduce the variants and analyze the component effects respectively:

(1). \textit{Remove \textbf{C}luster-wise \textbf{R}e-weighting (w/o CR)}: The most significant debiasing ability and overall accuracy degradation occurred without the re-weighting strategy. This proves the effectiveness of our proposed method: clustering the items and treating the recommendation target as an item cluster-wise multi-objective optimization problem can greatly alleviate the inherent bias issues of the model and improve its performance.

(2). \textit{Remove \textbf{P}opularity \textbf{D}iscrepancy-based Bisecting Clustering (w/o PD)}:  
In this variant, we replace the item clustering method with  \textit{AE-KMeans}. After ignoring the item popularity factor and performing clustering only in the item representation space, we find that the overall performance gets dropped. This result validates that item popularity should be a necessarily concerned during its clustering process and our popularity discrepancy-based clustering method can help distinguish different item cluster-wise optimization objectives.


(3) \textit{Remove \textbf{C}ounterfactual \textbf{I}nference (w/o CI)}: If the user global propensity causal path is kept, we find that the debiasing metrics get harmed heavily, indicating that using the combined causal model for inference is biased by the item popularity. On the other hand, with the participation of counterfactual inference, we perform inference purely based on the user's individual interest, which can better reflect the user’s true preference.


To sum up, the combination of the above strategies (i.e., ICPE) yields the best performance, proving that all the components of ICPE are effective and work collaboratively to improve the recommendation debiasing performance and overall accuracy.

\begin{table}[t]
    \large
    \renewcommand\tabcolsep{4.0pt}
  \caption{Ablation study on three datasets. Boldface denotes the highest scores. w/o denotes without. $\downarrow$ indicates a severe performance drop (more than 10\%).}
  \label{ablation}
\resizebox{0.50\textwidth}{0.132\textwidth}{%
  \begin{tabular}{llllllll}
    \hline
    \multirow{2}*{Dataset} & \multirow{2}*{Variant} & \multicolumn{4}{c}{Debiasing Metrics} & \multicolumn{2}{c}{Overall Metrics} \\
    & &$RT@20$ &$NT@20$  &$Cov@20$ &$AT@20$ &$R@20$ &$NG@20$  \\
    \hline
    \multirow{4}*{Last.Fm} 
    &w/o CR  &  \ 0.0061$\downarrow$ &\ 	0.0036$\downarrow$ &	\ 0.4248 &\ 	0.1869 &\ 	0.0414  &\ 	0.0391       \\ 
    &w/o PD  &\   0.0064 &\ 	0.0040  &\ 	0.4373 &\ 	0.1925 &	\ 0.0432 &\ 	0.0399        \\ 
	&w/o CI &\   0.0064 &\ 	0.0038$\downarrow$  &\ 	0.4340 &\ 	0.1890 &\ 	0.0420 &\ 	0.0395         \\ 
	\cline{2-8}
	&Default  & \  \textbf{0.0068}  &  \  \textbf{0.0044}   &\  \textbf{0.4418} &\  \textbf{0.2043} &\  \textbf{0.0432}    &\  \textbf{0.0404}      \\ 
    \hline
    \multirow{4}*{Gowalla} &w/o CR    &\  0.0799$\downarrow$ &\ 	0.0356$\downarrow$  &\ 	0.4034$\downarrow$ &\ 	0.1428$\downarrow$ &	\ 0.1784 &\ 	0.1557      \\ 
    &w/o PD    &\  0.0857 &\ 	0.0382  &\ 	0.4368$\downarrow$ &\ 	0.1573 &\ 	0.1845 &\ 	0.1586        \\ 
	&w/o CI  &\  0.0843 &\ 	0.0367  &\  0.4277$\downarrow$ &\ 	0.1508 &\ 	0.1837 &\  	0.1574         \\
	\cline{2-8}
	&Default  &\  \textbf{0.0897}  & \ \textbf{0.0403}    &\  \textbf{0.4860} &\  \textbf{0.1600} &\  \textbf{0.1898}    &\  \textbf{0.1626}      \\ 
    \hline
    \multirow{4}*{Yelp2018} &w/o CR    &\  0.0159$\downarrow$ &\ 	0.0080$\downarrow$ &\ 	0.4011$\downarrow$ &\  	0.0945$\downarrow$  &\ 	0.0550 &\ 	0.0421      \\ 
    &w/o PD    &\  0.0171 &\ 	0.0091  &\ 	0.4359 &\ 	0.1119 &\ 	0.0576 &\ 	0.0427         \\ 
	&w/o CI   &\  0.0169 &\ 	0.0085  &\ 	0.4227 &\ 	0.1112 &\ 	0.0574 &\ 	0.0424         \\ 
	\cline{2-8}
	&Default  &\  \textbf{0.0177}  &\  \textbf{0.0093}    &\  \textbf{0.4677} &\  \textbf{0.1168}  &\  \textbf{0.0599}    &\  \textbf{0.0448}      \\ 
    \hline
  \end{tabular}
  }
  
\end{table}

\begin{figure} [t]
  \includegraphics[width=0.5\textwidth]{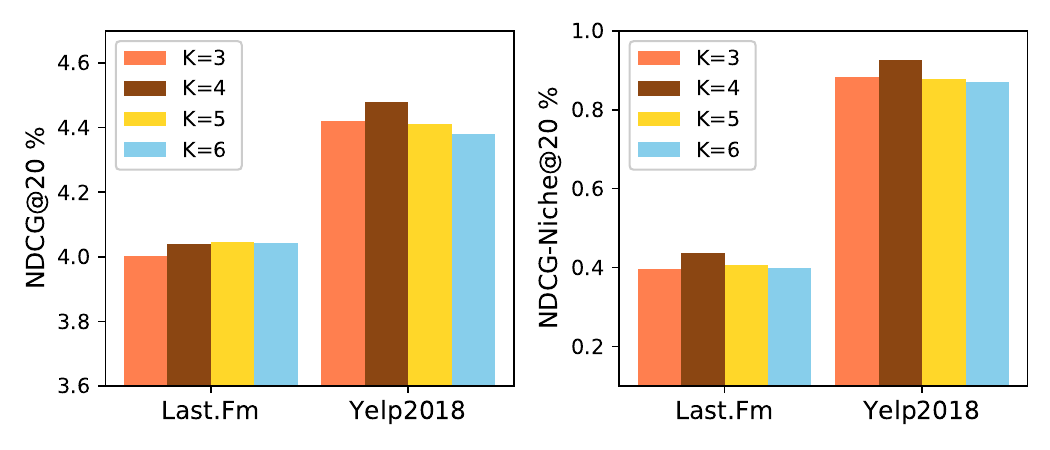}
  \caption{Effect of item cluster number $K$ in ICPE.}
  \vspace{-0.5cm}
    \label{hyper_K}
\end{figure}

\subsubsection{Hyper-parameter Study} 

\indent (1). \textit{Effect of Item Cluster Number $K$}: In this part, we use LightGCN as the backbone recommender model since it has the start-of-the-art overall accuracy performance.  
Here we choose $K \in \{ 3, 4, 5, 6 \}$ to conduct experiment.
Fig. \ref{hyper_K} illustrates, NDCG@20 and NDCG-Niche@20 under different cluster numbers on Last.Fm and Yelp2018 dataset. A similar trend can also be observed on other datasets and backbones. The general observation is that overall recommendation accuracy basically maintains the same level while the debiasing performance and accuracy on the niche item set manifest bell-shaped curves. Increasing the cluster number from 3 to 4 leads to the largest performance gain. Later on, the model's debiasing ability and performance on the niche item set keep diminishing along with the increase of cluster number. 
Such experimental results indicate that adaptively assigning the items into $4$ clusters leads to the most 
satisfactory performance. However, with more item clusters, the ability of ICPE in debiasing gets compromised. The reason could be that ICPE would put more focus on the balance between clusters that differ in representation information rather than the balance between head and niche items when $K$ is too large.

\begin{figure} [t]
\hspace*{-0.4cm}
  \includegraphics[width=0.5\textwidth]{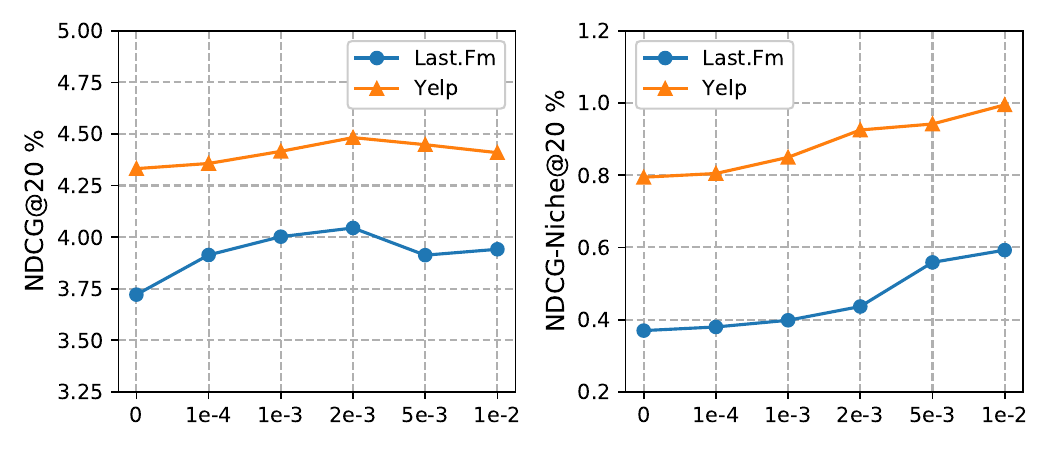}
  \caption{Effect of User Global Propensity Causal Path Coefficient $\alpha$.}
    \label{hyper_lambda_p}
\end{figure}

(2). \textit{Effect of User Global Propensity Causal Path Coefficient $\alpha$}:
To evaluate the impact of coefficient $\alpha$, we vary its value in the range of $\{0, 1 \times 10^{-4}, 1 \times 10^{-3}, 2 \times 10^{-3}, 5 \times 10^{-3}, 1 \times 10^{-2}\}$. The experimental results are summarized in Fig. \ref{hyper_lambda_p}. We can observe that the overall NDCG@20 reaches its peak when $\alpha = 2 \times 10^{-3}$, thus demonstrating that properly decomposing the factors behind user interactions can truly extract users' individual interest. Meanwhile, with the increment of $\alpha$, the model's debiasing ability and accuracy on the niche item set keep rising swiftly. This implies that more niche items are excavated when we put more emphasis on 
the user global propensity. To sum up, for higher overall accuracy, we choose $\alpha = 2 \times 10^{-3}$ as our default setting.


\subsection{Investigation of Item Cluster-wise Objective Optimization Solution  (RQ3)}
In this part, we conduct experiments to demonstrate whether ICPE generates a debiased Pareto-efficient solution and attaches proper importance to the niche item set.



\begin{figure}[t]
\centering    
\subfigure[Pareto Frontier and the found solutions] {
 \label{frontier}     
\includegraphics[width=0.23\textwidth]{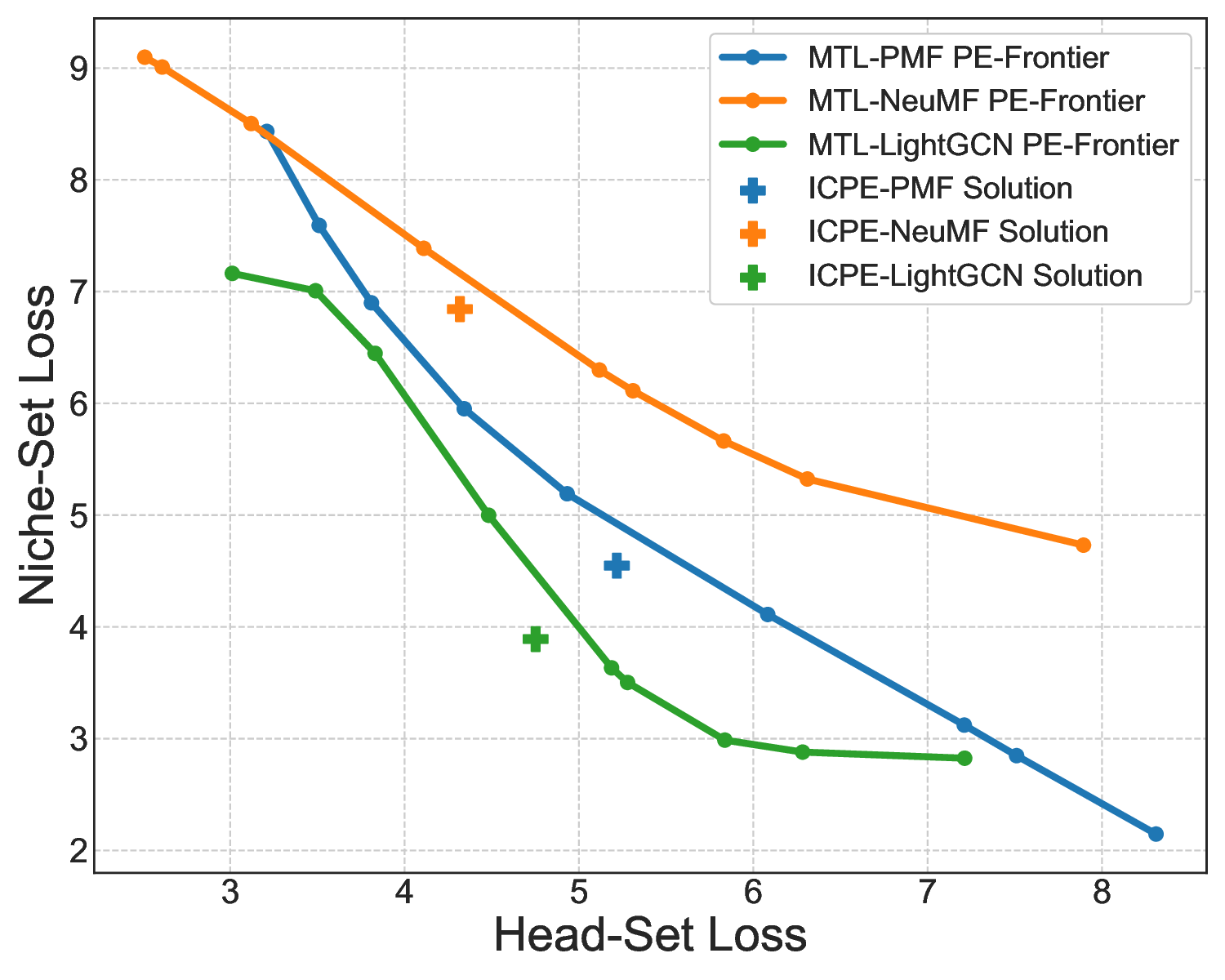}  
} 
\subfigure[Niche set weights learning curves] {
\label{lightgcn_weights}     
\includegraphics[width=0.23\textwidth]{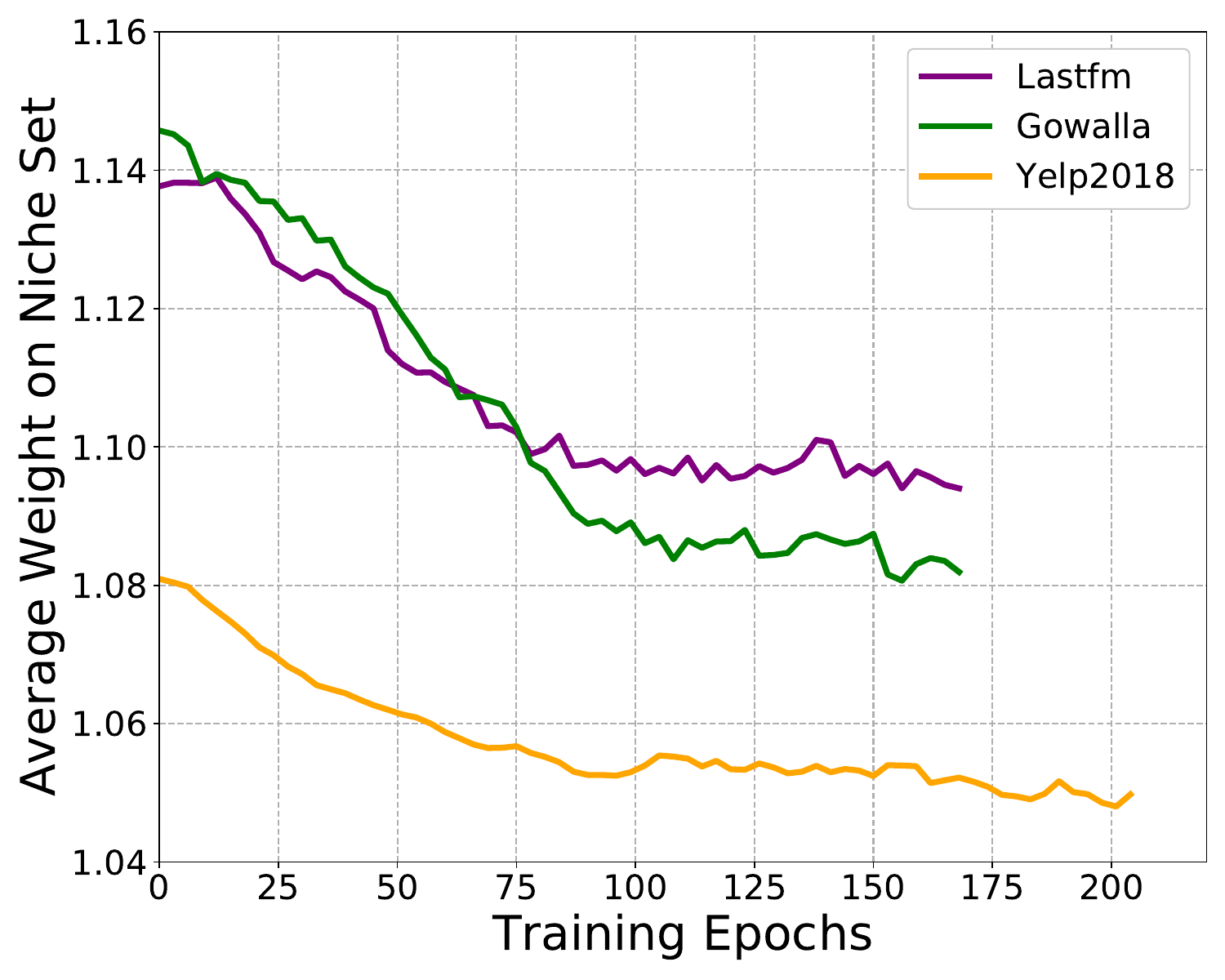}  
}     
\caption{Investigation of Item Cluster-wise Objective Optimization Solution.}     
\label{pareto} 
\vspace{-0.35cm}
\end{figure}

\subsubsection{Pareto Frontier and the Searched PE Point} 
On the Gowalla dataset with all the three recommender models, we mark our ICPE solution and generate the Pareto Frontiers of head-set and niche-set losses by running the Pareto MTL algorithm \cite{lin2019pareto} with various trade-off preference vectors, shown in Fig. \ref{frontier}. 
It can be observed that the obtained Pareto Frontiers under different constraints follow Pareto-efficiency, i.e., no point achieves both lower head and niche losses than other points. 
When the model focuses more on the head item set, the head-set loss is lower while the niche-set loss increases, and vice versa.

When it comes to the found solution of ICPE, we can observe that on all recommender models, those points mainly lie in the middle part of the Pareto Frontiers. 
This observation indicates that the \textbf{Pareto-efficiency solver} component of ICPE coincides with our aim of balancing the trade-off between head items and niche items. Furthermore, the solutions of ICPE obtain fewer losses on both of the two objectives than the Pareto Frontier generated by Pareto MTL, manifesting its superiority.





\begin{figure} [t]
  \centering
  \includegraphics[width=0.475\textwidth]{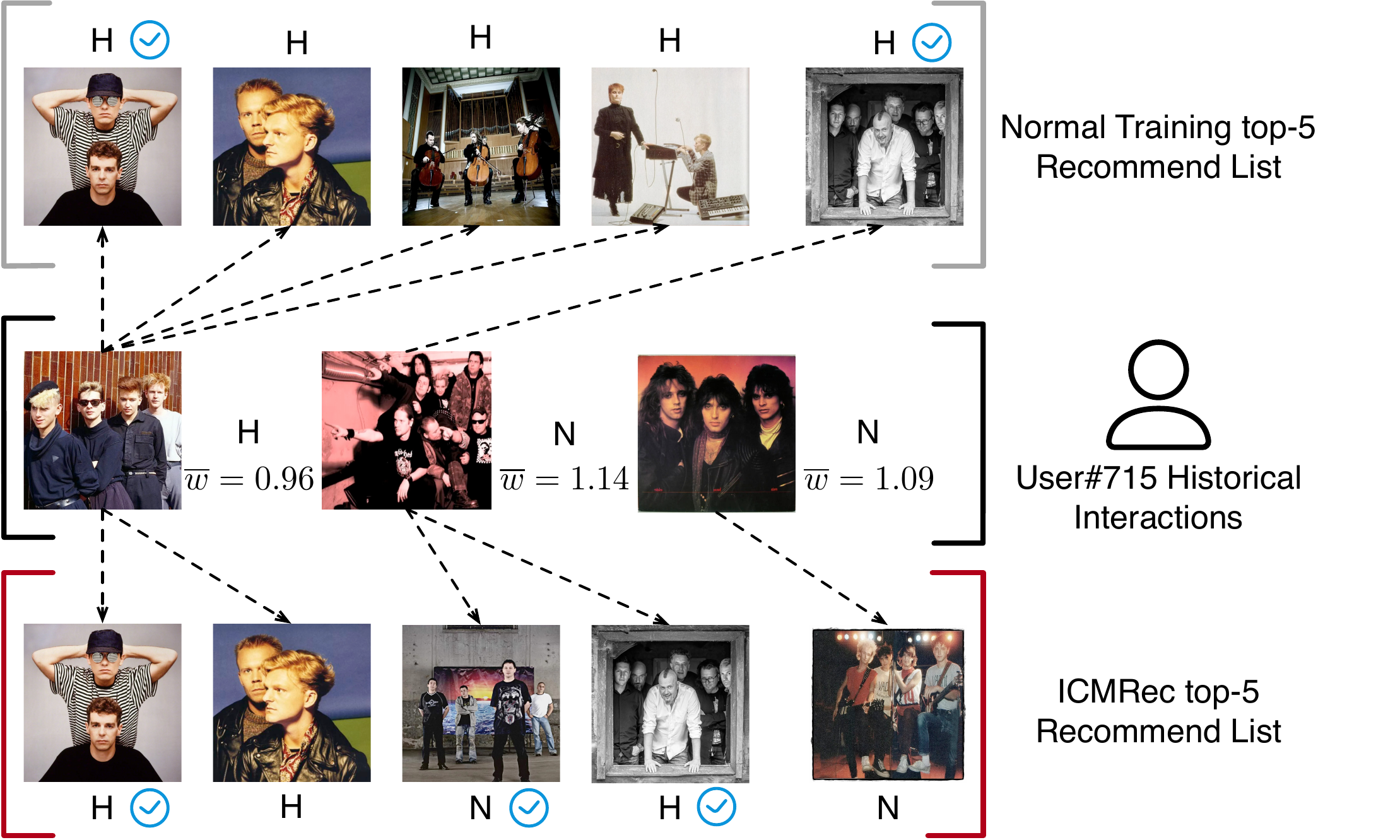}
  \caption{Case study of a single user. H denotes head item while N denotes niche item. $\overline{w}$ denotes the average weight assigned to each item. The arrow indicates that there exist relations between the pair of items (e.g., same tag).}
\label{case_study}
  \vspace{-0.35cm}
\end{figure}

\begin{table}
\begin{center}
\renewcommand\arraystretch{1.2}
  \caption{Case study of averaged interaction number of items in each cluster of ICPE (in descending order, $K = 4$).}
  \vspace{-0.1cm}
  \label{tab:cluster_numbers}
  \begin{tabular}{ccccc}
    \hline
    DataSet & $\mathcal{N}_{C_1}$ & $\mathcal{N}_{C_2}$ & $\mathcal{N}_{C_3}$ &  $\mathcal{N}_{C_4}$ \\
    \hline
    \textbf{Last.Fm}& 74.34 & 12.33 & 9.25 & 3.61 \\ 
    \hline
    \textbf{Gowalla}& 65.31 & 15.86 & 10.12 & 2.95 \\
    \hline
    \textbf{Yelp2018} & 97.91 & 18.62 & 11.22  & 2.60 \\
    \hline
  \end{tabular}
  \vspace{-0.3cm}
\end{center}
\end{table}

\subsubsection{Weights Learning Curves}
To be clear of the training process and weight assignment, we further plot the curves of the average weights assigned to niche items until convergence, as shown in Fig. \ref{lightgcn_weights}. We use LightGCN as the backbone recommender model on all three datasets. We can observe that the found weight solutions of ICPE mainly focus on the niche item set. After variating at the early training stage, the weight for the niche set becomes flattened and then converges to a value around $[1.04, 1.12]$. The reason draws on the fact that the gradient norm divergence between head and niche items is highly large  at the initial training stage; later on, when the model parameters have been updated towards the gradient direction of head items, the gradients' norm of head items reduced while those of niche items get amplified in scale.
On the other hand, the normal training setting neglects these Pareto-efficient weights and treats all items in a unified optimization target, leading to the bias toward head items and pre-exposed items. 



\subsection{Case Study (RQ4)}
To interpret the effectiveness of ICPE, on the Last.Fm dataset, we randomly select a user $u_{715}$ and retrieve his two top-5 recommendation lists from LightGCN-Normal and LightGCN-ICPE given the same historical interactions, visualized in
Fig. \ref{case_study}.
We observe that the recommendation list derived from LightGCN-Normal contains popular ground-truth items but does not contain any niche items. 
Due to the proper weights assigned to niche items in his interaction history, LightGCN-ICPE contains two niche items. 
One of the two recommended niche items belongs to the ground-truth test set, which improves the model's debiasing ability and overall accuracy. Note that the two recommendation lists have several items in common, which indicates that LightGCN-ICPE can also capture users' preferences for head items.

Given the first training step at epoch $100$, we list the average interaction number of items in each cluster of the Popularity Discrepancy-based Clustering in Table \ref{tab:cluster_numbers}.  It is noteworthy that the averaged interaction numbers of each cluster exhibit significant variations, demonstrating the effectiveness of representing the popularity level with the global propensity.

 \section{RELATED WORK}
\label{related_work}


\subsection{Methods for Recommendation Debiasing}

Inverse Propensity Weighting (IPW) is a widely-used debiasing method for RS, which re-weights each sample loss to recover the unbiased distribution. AutoDebias \cite{chen2021autodebias} combines IPW with a meta-learning paradigm to find a universal solution for debiasing. Jae-woong et al. \cite{lee2022bilateral} designed BISER that reuses the model’s prior prediction as the sample’s propensity score of the current training step. Nonetheless, the high variance of estimated propensity score is the main drawback of these IPW-based methods. By contrast, in the cluster-wise optimization setting of ICPE, such an issue has been greatly mitigated. 

Based on causal graphs and operators, causal reasoning methods \cite{he2022causpref} could also eliminate the negative effect of item popularity in recommendation. DICE \cite{zheng2021disentangling} assigns two independent representations to each user and item, respectively modeling the separate effects of relevance and conformity. However, assigning each item a unique conformity embedding may lead to model over-parameterization and overfitting. MACR \cite{wei2021model} removes the direct causal effect of user conformity and item conformity through counterfactual inference alone. However, the above methods could not address the exposure bias caused by the self-loop of RS (Figure \ref{self_loop}).

Besides, some empirically designed methods relieve the bias factors based on domain knowledge input such as side-information, prior hypothesis \cite{zhu2021popularity}, and niche item clustering \cite{bai2017dltsr,kim2019sequential}; CPR \cite{wan2022cross} forming the unbiased loss term as the combination of multiple observed interactions at once; Some works \cite{liu2020general,chen2020esam} introduce domain adaptation by gaining domain knowledge from a small but unbiased dataset and apply it to the main-biased dataset for training. 

In summary, none of the existing methods emphasized solving the bias issues in RS during the training process from item gradient or cluster-wise objective optimization perspective. In this work, we define recommendation as an item cluster-wise objective optimization problem, guiding the model to balance the gradient update on all item clusters that differ in popularity. 


\subsection{Multi-Objective Optimization in Recommendation}
In the research field of RS, despite that overall accuracy is always set as the main objective of recommendation, some researches have also been done focusing on other objectives such as availability, profitability, and usefulness \cite{jambor2010optimizing, jin2023predicting, jin2023code}.
Besides, metrics about diversified recommendations such as diversity, novelty, and fairness are also considered as objectives \cite{ribeiro2014multiobjective}. Recently, user-oriented objectives such as user sentiment are considered for better recommendation \cite{musto2017multi}. 
As for a commercial RS, CTR (Click Through Rate) and GMV (Gross Merchandise Volume) are included in the objectives \cite{nguyen2017multi} to gain higher profits.

The optimization methods for multiple objectives 
can be categorized into two categories: heuristic search and scalarization \cite{ xiao2017fairness}. Evolutionary algorithms are popular choices for heuristic search which
deal simultaneously with a series of possible solutions in a single run. But these algorithms usually depend heavily on the heuristic experience \cite{zitzler2001spea2}.
Scalarization methods transform multiple objectives into a single one with a weighted sum of all objective functions \cite{xiao2017fairness}. Then the overall objective function is optimized to be Pareto-efficient, where no single objective can
be further improved without hurting the others \cite{lin2019pareto}.

Unlike existing methods which introduce new metrics as the objectives \cite{lin2019pareto}, we consider the learning on each cluster of items as an objective. After that, we focus on finding an optimal weight solution for the cluster-wise objectives so that an overall harmonious gradient direction can be obtained.

\section{Conclusion}
\label{conclusion}

In this paper, we propose to tackle the bias issues in recommendation from a multi-objective optimization perspective. We first find that head items are repetitively recommended due to the fact that head items tend to have larger gradient norms and thus dominate the gradient updates. To alleviate such a phenomenon, we propose a model-agnostic framework namely ICPE, modeling the recommendation task as an item cluster-wise multi-objective optimization problem. Specifically, ICPE is featured with popularity discrepancy-based clustering, Pareto-efficiency solver, and counterfactual inference. We instantiate ICPE with three state-of-the-art recommender models and conduct extensive experiments on three real-world datasets. The results demonstrate the effectiveness of ICPE. Future work includes generalizing ICPE for other similar tasks such as multi-class classification and long-tail document retrieval.


%

%
%
%
%
%

\ifCLASSOPTIONcaptionsoff
  \newpage
\fi



%

\bibliographystyle{IEEEtran}

\bibliography{ICMRec.bib}

%
%

%
\begin{IEEEbiography}[{\includegraphics[width=1in,height=1.25in,clip,keepaspectratio]{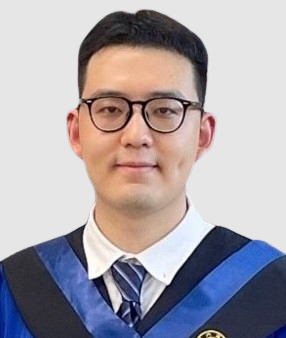}}]{Yule Wang} is currently working towards the Ph.D. degree in the School of Computational Science and Engineering, Georgia Institute of Technology (Gatech), United States. He received his M.S. and B.S. degree at Shanghai Jiao Tong University (SJTU). His major research interests include computational neuroscience, latent variable models and data mining.
\end{IEEEbiography}


\begin{IEEEbiography}[{\includegraphics[width=1in,height=1.25in,clip,keepaspectratio]{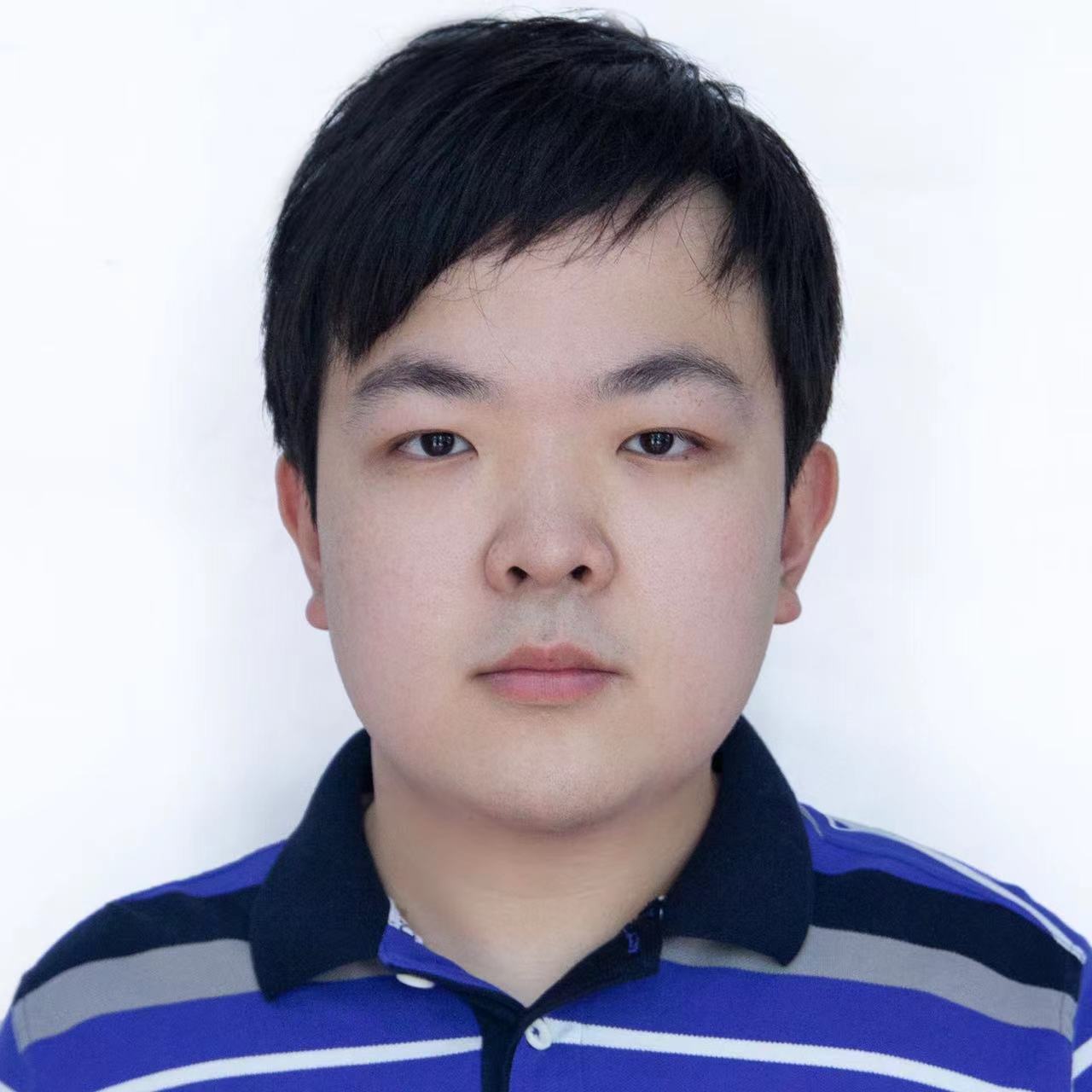}}]{Xin Xin} is now a tenure-track assistant professor in the school of computer science and technology, Shandong University. His research interests include machining learning and reinforcement learning for recommender systems and information retrieval. He has published more than 15 papers in top-ranking conferences, including SIGIR, IJCAI, ACL, WSDM, ect. He also serves as the program committee member and invited reviewer for tire-1 conferences and journals, such as SIGIR, IJCAI, WSDM, ACL, ACM Multimedia, TKDE, TOIS.
\end{IEEEbiography}

\begin{IEEEbiography}[{\includegraphics[width=1in,height=1.25in,clip,keepaspectratio]{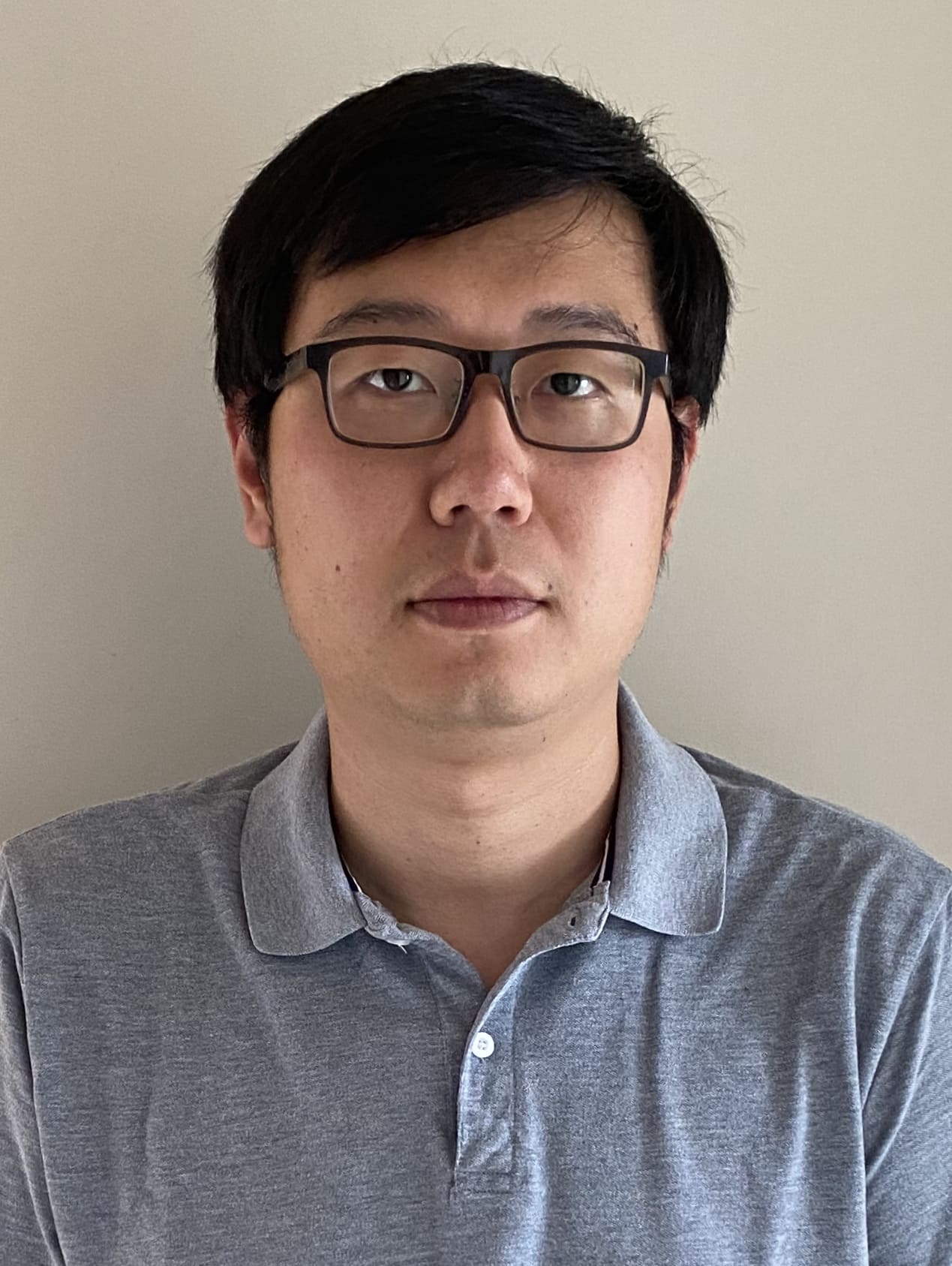}}]{Yue Ding} received the Ph.D. from Shanghai Jiao Tong University in 2018. He is now a lecturer in School of Software, Shanghai Jiao Tong University. His main research interests are recommender systems, deep learning, graph neural networks, and data mining.
\end{IEEEbiography}

\begin{IEEEbiography}[{\includegraphics[width=1in,height=1.25in,clip,keepaspectratio]{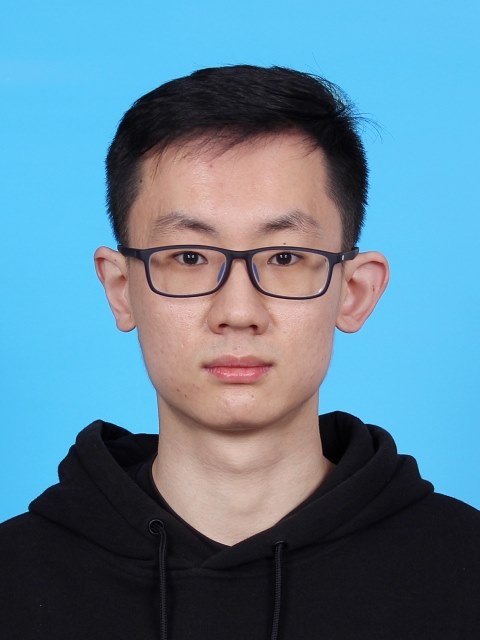}}]{Yunzhe Li} is a Ph.D. student in Department of Computer Science, University of Illinois, Urbana-Champaign(UIUC). His research interests involve information retrieval, natural language generation and recommender system.\end{IEEEbiography}

\begin{IEEEbiography}[{\includegraphics[width=1in,height=1.25in,clip,keepaspectratio]{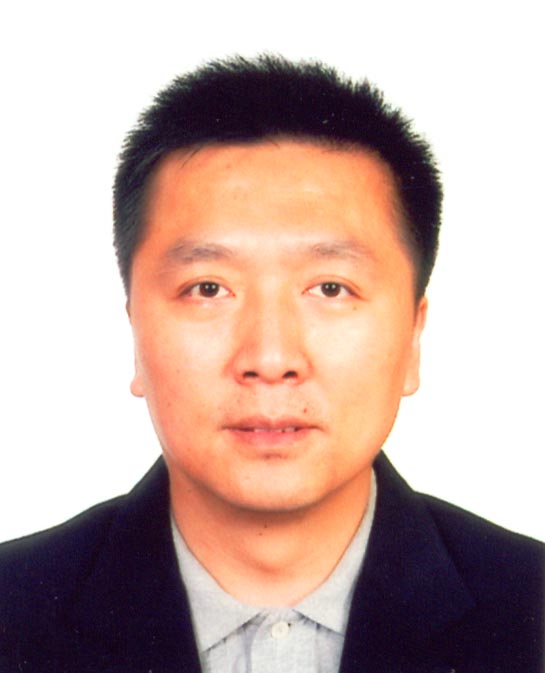}}]{Dong Wang} received the bachelor's and master's degrees from Xi'an Jiaotong University and Chongqing University, in 1991 and 1994, respectively, and the Ph.D. degree in mechanical manufacturing and automation from Shanghai Jiao Tong University (SJTU), in 2001. He was in charge of  many projects supported by programs like the National High Technology Research and Development Program. His main research interests include RFID, Internet of Things, pervasive computing, wireless sensing, data mining, and recommender systems. He is currently a professor with the School of Software, SJTU. He has many publications in top journals and conferences, like Ubicomp/IMWUT, Sensys, WWW, IJCAI, Percom and CIKM.\end{IEEEbiography}





\end{document}